\documentclass[journal]{IEEEtran}
\usepackage{bm}
\usepackage{mathrsfs}
\usepackage{graphicx}
\usepackage{cite}
\usepackage{color}
\usepackage{amsmath}
\usepackage[noend]{algpseudocode}
\usepackage{algorithmicx,algorithm}
\usepackage{subfigure}
\usepackage{xpatch}

\makeatother

\begin{document}
\pdfoutput=1
% paper title
% Titles are generally capitalized except for words such as a, an, and, as,
% at, but, by, for, in, nor, of, on, or, the, to and up, which are usually
% not capitalized unless they are the first or last word of the title.
% Linebreaks \\ can be used within to get better formatting as desired.
% Do not put math or special symbols in the title.
\title{DeepReceiver: A Deep Learning-Based Intelligent Receiver for Wireless Communications in the Physical Layer
\thanks{This work was supported in part by the National Natural Science Foundation of China under Grant U19B2015. \emph{(Corresponding author: Xiaoniu Yang)} }
\thanks{The authors are with Science and Technology on Communication Information Security Control Laboratory, Jiaxing 314033, China (e-mail: lianshizheng@126.com, sicanier@sina.com, yxn2117@126.com).}}
\author{Shilian~Zheng,
        Shichuan~Chen,
        and Xiaoniu~Yang
        % <-this % stops a space
}

\maketitle

% As a general rule, do not put math, special symbols or citations
% in the abstract or keywords.
\begin{abstract}
A canonical wireless communication system consists of a transmitter and a receiver. The information bit stream is transmitted after coding, modulation, and pulse shaping. Due to the effects of radio frequency (RF) impairments, channel fading, noise and interference, the signal arriving at the receiver will be distorted. The receiver needs to recover the original information from the distorted signal. In this paper, we propose a new receiver model, namely DeepReceiver, that uses a deep neural network to replace the traditional receiver's entire information recovery process. We design a one-dimensional convolution DenseNet (1D-Conv-DenseNet) structure, in which global pooling is used to improve the adaptability of the network to different input signal lengths. Multiple binary classifiers are used at the final classification layer to achieve multi-bit information stream recovery. We also exploit the DeepReceiver for unified blind reception of multiple modulation and coding schemes (MCSs) by including signal samples of corresponding MCSs in the training set. Simulation results show that the proposed DeepReceiver performs better than traditional step-by-step serial hard decision receiver in terms of bit error rate under the influence of various factors such as noise, RF impairments, multipath fading, cochannel interference, dynamic environment, and unified reception of multiple MCSs.
\end{abstract}
% Note that keywords are not normally used for peerreview papers.

\IEEEpeerreviewmaketitle
\section{Introduction}
\subsection{Background and Motivation}
\IEEEPARstart{W}{ith}  the large-scale deployment of 5G networks \cite{1}\cite{2} and the emergence of the Internet of Things \cite{3}, wireless communication has become an indispensable way for human society to communicate information. The physical layer wireless communication system usually consists of a transmitter and a receiver. The transmitter performs source/channel coding, modulation, and pulse shaping on the information to be transmitted, and then radiates the signal into the air through an antenna. In real-world communication process, when the signal reaches the receiver, it will be affected by various non-ideal factors, including radio frequency (RF) impairments, channel fading, noise, and interference. Among them, RF impairments are mainly caused by non-ideal characteristics of RF devices, which will cause carrier frequency deviation and/or in-phase and quadrature (IQ) imbalance in the received signal. Channel fading is mainly caused by geographical environment and obstructions, and usually includes shadowing and multipath fading, which will introduce inter symbol interference (ISI) in the received signal. Noise includes atmospheric thermal noise, industrial noise, and the system's noise which will also affect the quality of the received signal. Interference refers to unintentional or malicious interference from other emitters. When the receiver lacks specific anti-interference measures, the quality of information recovery will be seriously affected.

Traditional physical layer receivers often use processes such as carrier and symbol synchronization, channel estimation, equalization, demodulation, and decoding to recover information from the received distorted signals as accurately as possible. There are three limitations for this approach. First, the step-by-step serial processing does not optimize the overall performance of the receiver. Each module, such as carrier and symbol synchronization, channel estimation, equalization, demodulation, or decoding, optimizes the performance of the specific task. However, the optimal local performance of each module does not necessarily guarantee the optimal global performance. The errors of the pre-processing module may affect the optimization of the subsequent processing modules, resulting in the cumulative effect of errors. Second, the algorithm design of each processing module is usually based on theoretical assumptions, such as assuming that the RF devices are ideal, the channel fading follows the Rice model, the noise is additive white Gaussian noise (AWGN), or there is no co-channel interference. These assumptions do not necessarily match the real conditions experienced by the communication system. Therefore, what the traditional receiver optimizes is the best performance under the assumptions, and not necessarily the best performance under the real-world environment. Finally, with the development of software radio technology, technologies such as adaptive modulation and coding (AMC) are commonly used. Traditional reception algorithms are often designed for specific modulation and coding. The receiver needs to know which modulation and coding scheme (MCS) the transmitter adopts to realize information recovery. This increases the complexity of the signaling interaction between the transmitter and receiver.

In order to cope with the above-mentioned problems of the traditional physical layer receiver, in this paper, we introduce the deep learning (DL) technology \cite{4} currently widely used in the fields of image, speech, natural language and other fields to the design of communication receiver. The reason why deep learning is adopted is because it is an end-to-end learning method and it can learn deeper features from large amount of data compared with other machine learning methods \cite{new5,new6,new7}. We propose a new receiver model, namely DeepReceiver, replacing the traditional receiver's information recovery process with a deep neural network model. The input of this model is the received IQ signal and the output is the recovered information bit stream. The model is trained based on the received IQ signal samples, and it is more able to reflect the RF impairments, channel fading, noise, and interference that the communication system actually experiences. Moreover, the same DeepReceiver model can be used for blind reception of multiple MCSs. We will comprehensively analyze and verify the performance of DeepReceiver under the conditions of noise, RF impairments, channel fading, cochannel interference, dynamic environment, and blind reception through extensive simulation experiments.
\subsection{Related Work}
With the development of deep learning technology, there are an increasing number of studies using deep learning for physical layer communication receivers. The following provides an overview of related work from three aspects.

\emph{1) Using DL to improve the performance of a specific processing module of the communication receiver:} In terms of channel estimation, in \cite{5} a DL-based channel estimator under time-varying Rayleigh fading channel was proposed and the proposed DL-based estimator has better Mean Square Error (MSE) performance compared with the traditional algorithms. In \cite{6}, a DL-based channel estimator was first trained offline using simulated data and then dynamically adjusted online to effectively improve the generalization ability of the estimator. There are also studies that used deep learning to address the problems of orthogonal frequency division multiplexing (OFDM) channel estimation \cite{7}\cite{8} and multiple-input multiple-output (MIMO) channel estimation \cite{9,10,11}.

In terms of channel equalization, in \cite{12} deep learning was used for channel equalization and several deep neural network-based equalizers were presented and compared with traditional equalization methods. Better error performance than traditional equalization methods was obtained.

In terms of signal demodulation, deep belief networks and stacked autoencoders were used in \cite{13} to complete signal demodulation in short-distance multipath channels. Convolutional neural network (CNN) was used to demodulate bipolar extended binary phase shifting keying (BPSK) signals transmitted at a faster-than Nyquist rate \cite{14}, solving the problem of severe ISI. In addition to hard demodulation, deep neural networks were also used to implement soft demodulation \cite{15}, which reduces the computational complexity and improves the demodulation performance.

In channel decoding, there are many studies that use deep learning in combination with belief propagation algorithms to improve decoding performance. Deep neural networks (DNNs) \cite{16}\cite{17}, recurrent neural networks (RNNs) \cite{18}, CNN \cite{19}, graph convolutional network (GCN) \cite{20} and other neural network structures were used. In view of the high computational complexity of the BP algorithm, in \cite{21}\cite{22} deep learning-based minimum sum decoding algorithms were proposed, which reduced the computational complexity and improved the decoding speed. In terms of Polar code decoding, there are also many works that use deep learning to improve the performance of Polar code decoding \cite{23,24,25,26}, reduce the delay of decoding \cite{27,28,29}, and facilitate hardware implementation \cite{30,31,32}.

In addition to replacing a certain processing module of the receiver with deep learning, some studies used deep learning to simultaneously optimize multiple processing modules of the receiver. For instance, in \cite{33} bidirectional recurrent neural network (BRNN) was used for data sequence detection, however channel decoding module was not included in the BRNN. In \cite{34} deep neural network was proposed to replace the equalization and decoding modules to improve the performance in multipath channels, but there is still a certain gap between the performance of minimal MSE method that knows the channel statistics. Based on this work, in \cite{35} deep neural network structure was used to replace the channel equalization and symbol detection modules in the OFDM receiver. Better performance than the traditional methods was obtained. In \cite{36}, two separate neural networks were used to replace the equalization module and decoding module, respectively. The two networks can be jointly trained to obtain better performance. In \cite{37}, a deep learning-based channel estimation and equalization method was used to address the problem of the traditional channel estimation algorithm in dealing with inherent interference. The pilot extraction, channel equalization, and quadrature amplitude modulation (QAM) demapping modules at the OFDM receiver were replaced with a deep neural network.

From the above discussion, it can be seen that the current applications of deep learning in the wireless communication receiver mainly focus on replacing one or several (but not all of the) modules in the receiver with a deep neural network. In addition, most of the current research on DL-based communication signal processing only considered the effects of wireless channel fading and there are almost no cases involving RF impairment, non-AWGN noise, channel interference and so on. However, the DeepReceiver model we propose in this paper will replace the entire link of information recovery and consider the impact of RF impairments, wireless channel fading, noise, and interference.

\emph{2) DL-based communications system:} In addition to using deep learning to learn the function of a certain module of the receiver, there are some research work on end-to-end optimization of wireless communication systems based on deep learning \cite{38,39,40,41,42}. These works used a deep neural network (commonly autoencoder) to replace the entire communication system, including the transmitter and the receiver. The neural networks at the transmitting and receiving ends are jointly optimized as a whole. This kind of DL-based communication system is a disruptive communication system. The DL-based receiving end can only be used for receiving and processing the signal generated by the corresponding DL-based transmitting end. It cannot be used to process signals generated by traditional communication transmitters. However, the DeepReceiver model we propose in this paper is a receiver model for information recovery of signals generated by traditional transmitters.

\emph{3) Unified receiver for multiple MCSs:} Traditional communication receiving algorithms are mainly designed for specific MCS. For a communication system that uses ACM at the transmitter end, the receiver end often needs to know which MCS the current signal uses to be able to select the corresponding information recovery algorithm. Another method is to first identify the modulation and coding of the received signal (where DL-based methods can be used \cite{new46,new47,new48,new50}), and then select the corresponding demodulation and decoding algorithm for information recovery. However, this method is still a serial processing method. In the case of low signal-to-noise ratio (SNR), the accuracy of modulation recognition and coding recognition may be greatly affected. Once the recognition error occurs, it will cause high demodulation and decoding error rate. The DeepReceiver we propose in this paper is essentially a unified blind processing method for multiple MCSs. During the training phase, the training set contains signal data of multiple MCSs. The trained DeepReceiver model can adapt to these MCSs and recover the information from the received signal without knowing in prior which MCS the transmitter adopts.

\subsection{Contributions and Structure of the Paper}
The contributions of the paper are mainly as follows:
\begin{itemize}
  \item[$\bullet$] We propose a DL-based DeepReceiver model that uses a deep neural network to recover the original information bit stream from the received distorted IQ signal. This model replaces the entire information recovery process of traditional receivers including carrier and symbol synchronization, channel estimation, equalization, demodulation, and decoding, trying to overcome the impact of RF impairments, wireless channel fading, noise, and cochannel interference as much as possible.
  \item[$\bullet$] We propose an implementation of DeepReceiver based on binary classifications sharing a common CNN. We specifically design a one-dimensional convolution DenseNet (1D-Conv-DenseNet) network structure, in which all convolutions are one-dimensional. Global pooling is used to obtain feature vectors of the same dimension to improve the adaptability of the network to different input signal lengths. In the final classification layer, multiple binary classifiers are used to recover the bit stream.
  \item[$\bullet$] We use DeepReceiver to realize the unified information recovery of multiple MCSs. We include signal samples generated with multiple MCSs in the training set, and the trained receiver model can recover the information bit stream without knowing the specific MCS adopted by the newly received signal.
  \item[$\bullet$] We evaluate the performance of DeepReceiver by extensive simulation experiments under several non-ideal factors including noise (AWGN and additive generalized Gaussian noise (AGGN)), RF impairments (carrier frequency deviation and IQ imbalance), multipath fading (frequency flat Rayleigh fading and frequency selective Rayleigh fading), cochannel interference (single-tone interference, minimum shift keying (MSK) interference and BPSK interference) and dynamic environment. Results show that DeepReceiver has superior performance under these circumstances. Especially the performance under the cochannel interference shows that the DeepReceiver is expected to become a new anti-jamming communications scheme.

\end{itemize}

The rest of the paper is organized as follows. Section II introduces briefly the Basic Wireless Communication System. Section III discusses the proposed DeepReceiver in detail. Section IV gives the simulation results in various cases and analyzes the performance of the DeepReceiver in detail. Finally, Section V concludes the paper and gives some orientations for future work.

\section{Brief Introduction of a Canonical Wireless Communication System}
\subsection{Brief Review of Traditional Receiver Model}
A canonical wireless communication system consists of a transmitter and a receiver, as shown in the upper part of Fig. 1. At the transmitter end, content such as voice, text, or video to be sent is converted into an information bit stream which can be denoted as ${\bm{s} = [{s_1},{s_2},...,{s_M}]}^T$ after being source encoded and encrypted, where ${[ \cdot ]^T}$ denotes transpose, ${s_m} \in \Theta  = \left\{ {0,1} \right\}$ and $M$ is the number of bits in the stream. The information bit stream is then channel encoded, modulated and pulsed shaped, after that, the resulting signal is radiated into the air by the antenna. When the receiver receives the signal, it adopts channel estimation, equalization, demodulation, and channel decoding to recover the information bit stream ${\bm{\widehat s} = [{\widehat s_1},{\widehat s_2},...,{\widehat s_M}]^T}$, and then decrypts and decodes the bit stream to get the original content.

\begin{figure*}[!t]
\centering
\includegraphics[width=13cm]{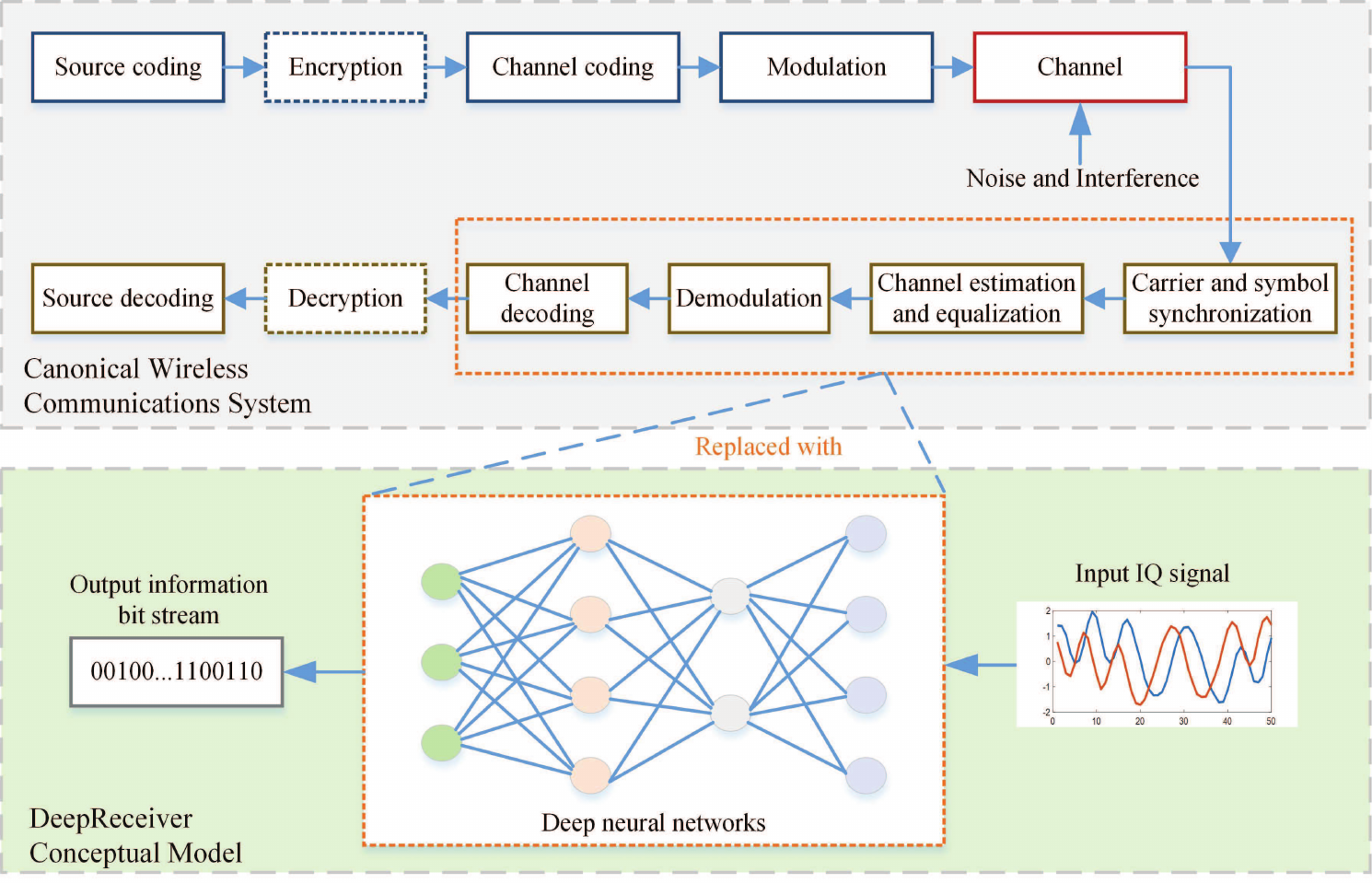}
% where an .eps filename suffix will be assumed under latex,
% and a .pdf suffix will be assumed for pdflatex; or what has been declared
% via \DeclareGraphicsExtensions.
\caption{Canonical wireless communication system and the proposed DeepReceiver conceptual model.}
\label{Fig.1}
\end{figure*}

The receiver is the key to ensuring the bit error rate (BER) performance of wireless communication systems. When the signal reaches the receiver, it will be affected by various non-ideal factors. The received signal can be represented as
\begin{equation}
r(t) = h\left( t \right) * x\left( t \right){e^{j(2\pi \Delta f + {\theta _0})t}} + n(t) + i(t),
\end{equation}
where $r(t)$ is the received signal, $x(t)$ is the transmitted pulse shaped signal, $h(t)$ is the channel pulse response, $\Delta f$ and $\theta _0$ are frequency and phase deviation respectively, $n(t)$ is the noise, and $i(t)$ is the interference. The receiver needs to recover the information from the distorted signal as accurately as possible. Assuming that the transmitted information sequence is uniformly distributed and the channel model is known, a maximum likelihood (ML) decision algorithm which is optimal in the sense of minimum probability of error is
\begin{equation}
\widehat {\bm{s}} = \mathop {\arg \max }\limits_{\bm{s} \in {\Theta ^M}} \Pr \left( {\left. {\bm{r}} \right|{\bm{s}}} \right),
\end{equation}
where $\Pr\left( {\left. {\bm{r}} \right|{\bm{s}}} \right)$ represents the likelihood probability, $\bm{r}={[r(0),r(1),...,r(N-1)]^T}$ is the digitalized received signal and $N$ is the length of the signal. The maximum likelihood decision needs to know the perfect channel model (including the noise and the interference), which is difficult to guarantee in practice. Instead, in order to overcome these non-ideal factors, traditional wireless communication receivers mainly use step-by-step serial processing to recover the information, i.e., using carrier synchronization to correct carrier frequency deviation, using symbol synchronization to overcome timing errors, using channel estimation to estimate the channel response, using equalization to overcome the channel fading, using demodulation to implement the inverse operation of the modulation, and using channel decoding for the inverse operation of the channel coding.

As pointed out earlier, in this receiving mode, the optimization of each module is the performance of the module itself, and not necessarily the overall global optimal performance of the information recovery of the communication system. The error of the pre-processing module may affect the optimization of the subsequent processing module, resulting in error accumulation. Furthermore, the algorithm design of each receiving processing module is usually based on theoretical assumptions, which may not necessarily match the non-ideal real conditions experienced by the communication system. Therefore, what the traditional receiver optimizes is the best performance under the assumptions, and not necessarily the best performance under the real-world environment. Finally, receiving algorithms are often designed for specific MCS, and it is difficult to adapt to the unified reception and information recovery of multiple MCSs.
\subsection{Non-ideal Factors}
In summary, we consider four factors that affect the quality of the received signal in this paper:
\begin{itemize}
\item[$\bullet$] \emph{RF impairments}. The difference between the transmitter and receiver local oscillators will cause the received signal to have frequency deviation. In addition, in actual hardware circuits, the physical limitations of the device and circuit design errors will cause the phase and amplitude of the I and Q signals to be inconsistent, resulting in IQ imbalance \cite{43}. The IQ imbalance can be represented as
      \begin{equation}
       \begin{array}{l}
y\left( x \right) = {\mathop{\rm Re}\nolimits} \left( x \right){10^{\frac{\alpha }{{40}}}}\exp \left\{ { - \frac{{j\beta \pi }}{{360}}} \right\}\\
\begin{array}{*{20}{c}}
{}&{}&{}
\end{array} + j{\mathop{\rm Im}\nolimits} \left( x \right){10^{\frac{\alpha }{{40}}}}\exp \left\{ { \frac{{j\beta \pi }}{{360}}} \right\},
\end{array}
      \end{equation}
      where ${\mathop{\rm Re}\nolimits} \left( x \right)$ and ${\mathop{\rm Im}\nolimits} \left( x \right)$ are real (I) and imaginary (Q) component of the signal respectively, $\alpha$ is the amplitude imbalance in dB and $\beta$ is the phase imbalance in degrees. We will analyze the performance of the algorithms in the cases of carrier frequency deviation and IQ imbalance in the simulation experiments.
\item[$\bullet$] \emph{Channel fading}. Terrain, obstacles and other factors may affect the propagation of the signal, leading to multipath fading with the received signal. In addition, the relative motion between the transmitter and receiver will cause the Doppler shift
     \begin{equation}
       \Delta f = \frac{{fv\cos \theta }}{{{v_c}}},
      \end{equation}
      where $f$ is the signal frequency, $v$ is the relative speed between the transmitter and the receiver, $\theta$ is the angle between the direction of motion and the incident direction of the signal, and $v_c$ is the speed of light. These factors will cause serious signal distortion such as ISI. In the simulation experiments, we will consider frequency flat Rayleigh fading channel and frequency selective Rayleigh fading channel and analyze the performance of the algorithms in these cases.
  \item[$\bullet$] \emph{Noise}. Due to the existence of atmospheric thermal noise and the noise of the communication system itself, the received signal will contain a certain amount of noise. The most common noise is AWGN. In addition, we also consider AGGN \cite{44} which can better characterize ``pulse'' noise. The probability density function of AGGN is
      \begin{equation}
      p(\omega ) = \frac{\rho }{{2\gamma \Gamma ({\raise0.7ex\hbox{$1$} \!\mathord{\left/
 {\vphantom {1 \rho }}\right.\kern-\nulldelimiterspace}
\!\lower0.7ex\hbox{$\rho $}})}}\exp \left\{ { - {{\left| {\frac{{\omega  - \mu }}{\gamma }} \right|}^\rho }} \right\},
\end{equation}
      where $\mu$ is the mean, $\rho$ is the ¡°shape parameter¡±, and $\Gamma \left(  \cdot  \right)$ is the Gamma function. When $\rho  = 2$, (5) becomes traditional Gaussian distribution. In the simulation experiments, we will analyze the performance of the algorithms under the two noise distributions of AWGN and AGGN.
  \item[$\bullet$] \emph{Interference}. In the electromagnetic spectrum space, signals from other emitters may cause co-channel interference to the receiver. When the power of the interference is large (relative to the received communication signal power), if there is no special anti-interference measure, the performance of the traditional communication receiver will be seriously deteriorated. In the simulation experiments, we will analyze the performance of the algorithms in the presence of co-channel single-tone interference, co-channel MSK interference and co-channel BPSK interference.
\end{itemize}

\section{The Proposed DeepReceiver Model}
\subsection{ Basic Concept of DeepReceiver}
In this paper, we coin \emph{DeepReceiver} as \emph{a unified, blind, and intelligent receiver based on a deep learning model (such as a CNN) which implements end-to-end information recovery from received communication signal waveform.} As shown in the lower part of Fig. 1, the DeepReceiver model uses a deep neural network to replace the information recovery process, including carrier and symbol synchronization, channel estimation, equalization, demodulation, and channel decoding. The input of the model is the received and sampled IQ signal ${\mathop{\rm Re}\nolimits} \left( {r\left( n \right)} \right)$ and ${\mathop{\rm Im}\nolimits} \left( {r\left( n \right)} \right)$ $(n=0,1,...,N-1)$, and the output is the recovered information bit stream $\bm{\widehat s} = {[{\widehat s_1},{\widehat s_2},...,{\widehat s_M}]^T}$, ${\widehat s_m} \in \Theta$. This DeepReceiver is used to recover information from signal transmitted by a traditional communication transmitter. Its purpose is to learn the complex relationship between the received signal and the transmitted information sequence and thus to reliably recover information under various non-ideal conditions as much as possible to improve the adaptability of the receiver to non-ideal conditions. With the goal of minimizing BER as an example, the optimization of DeepReceiver can be expressed as
\begin{equation}
\mathop {\min }\limits_{\mathcal {Q}} {\left\| {\bm{\widehat s}-\bm{s}} \right\|_1},\bm{\widehat s} = {\mathcal F}\left( {\left[ {{\mathop{\rm Re}\nolimits} \left( \bm{r} \right),{\mathop{\rm Im}\nolimits} \left( \bm{r} \right)} \right];{\cal {Q}}} \right),
\end{equation}
where $\mathcal{Q}$ represents the model parameters of the DeepReceiver, $\mathcal {F}\left( { \cdot ;{\mathcal Q}} \right)$ represents the function mapping of the DeepReceiver from the input to the output.

DeepReceiver has three main features. The first is global optimization. In the DeepReceiver model, a single deep neural network performs all processing of information recovery, and the network optimizes the overall performance of the information recovery. The second feature is that the DeepReceiver does not rely on theoretical assumptions. The DeepReceiver is designed based on deep learning which is a method of learning from data. The learned model will more closely match the non-ideal factors experienced by the communication system, and it is expected to obtain better performance than traditional receivers in these non-ideal situations. The third feature is its unity. The DeepReceiver can realize unified blind information recovery of multiple MCSs as long as the DeepReceiver has seen the signal samples of these MCSs during the training phase.

\subsection{CNN Classification-Based DeepReceiver Implementation}
As noted, let the transmitted information bit stream be $\bm{s} = {\left[ {{s_1},{s_2},...,{s_M}} \right]^T}$, where $M$ is the number of bits in the stream. The information bit stream is channel encoded, modulated, and pulse shaped. The resulting signal is radiated into the air and propagates through the wireless channel to the receiving end. The task of the DeepReceiver is to recover the information bit stream from the received IQ signal. The purpose is to make the recovered bit stream $\bm{\widehat s} = {[{\widehat s_1},{\widehat s_2},...,{\widehat s_M}]^T}$ as equal to the transmitted bit stream as possible. In this paper, we propose a DeepReceiver implementation based on CNN classification. The network structure is described in detail in the following.

\emph{1) Multiple binary classifiers:} The information recovery problem can be regarded as a sequence recognition problem, and one method is to solve it with a single multi-category classifier. The bit stream includes a total of $M$ bits, and the number of all possible classes is $2^M$, so a classifier with $2^M$ categories can be used to solve it. However, as the number of bits increases, the number of categories increases exponentially. For example, when $M = 32$, ${2^M} \approx 4.295 \times {10^9}$. A single classifier containing such a large number of categories is difficult to implement, mainly due to two reasons. Firstly, the number of hidden nodes in the final classification layer of the neural network is generally the same as the number of categories. The inclusion of such a large number of hidden nodes increases the time and space complexity of the network. Secondly, for each category, a certain number of training samples are often needed, so the number of training samples needed will be much larger than $2^M$. It is uneconomical to generate such a large number of training samples, and the computational complexity of training will become very high, which makes it difficult to converge in a limited time.

To solve this problem, we use $M$ binary classifiers at the final classification layer to recover $M$-bit information bit stream instead of a single multi-category classifier. Each binary classifier recovers one of the bits. The number of classifiers is consistent with the number of bits in the bit stream to be recovered. It should be noted that the $M$ binary classifiers we designed are not isolated. They share the same neural network. The overall structure is shown in Fig. 2. The digital sampled IQ signal $\left\{ {{\mathop{\rm Re}\nolimits} \left( {r\left( n \right)} \right),{\mathop{\rm Im}\nolimits} \left( {r\left( n \right)} \right)} \right\}\left( {n = 0,1,...,N - 1} \right)$ is used as the input of the CNN. After a series of operations of convolution, pooling, and activation, a feature vector is obtained. The feature vector is used as the input of the $M$ binary classifiers, and the outputs of the binary classifiers correspond to the recovered information bit stream. The specific structure of the designed CNN will be discussed in detail in the next section. It can be seen that the CNN is very important for DeepReceiver, which will determine the performance of the learned feature vector, thereby affecting the recovery performance of each bit.
\begin{figure}[!t]
\centering
\includegraphics[width=3.4in]{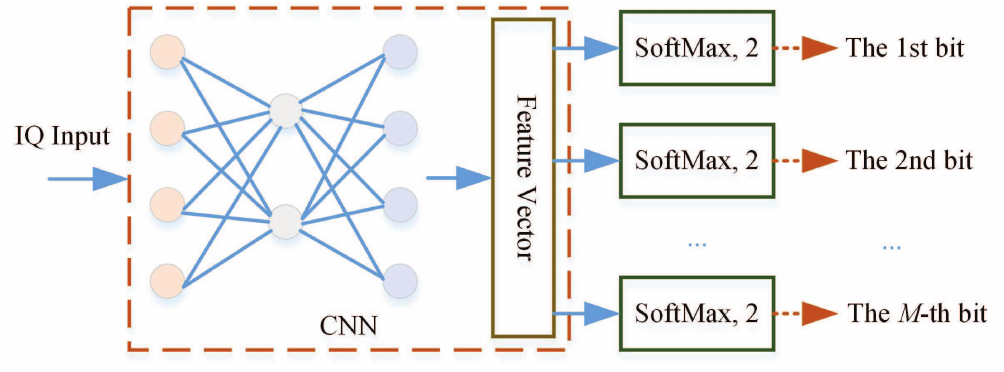}
% where an .eps filename suffix will be assumed under latex,
% and a .pdf suffix will be assumed for pdflatex; or what has been declared
% via \DeclareGraphicsExtensions.
\caption{DeepReceiver structure using a CNN with multiple binary classifiers.}
\label{Fig.2}
\end{figure}

\emph{2) The designed 1D-Conv-DenseNet structure:}
There are many excellent CNN structures for image classification, including Inception \cite{45}, ResNet \cite{46}, and DenseNet \cite{47}. Among them, DenseNet has achieved better performance than the other two networks. We will build our network for DeepReceiver based on DenseNet. Unlike traditional DenseNet used for image processing, we design 1D-Conv-DenseNet, in which the size of the convolution kernel of all convolutions is one-dimensional.

DenseNet is a densely connected network. In order to ensure the maximum information flow between the layers in the network, it directly connects all layers (with matching feature map sizes) to each other. Let $F_l (\cdot)$ be the $l$-th non-linear mapping layer in the network, and $\bm {y}_l$ be the output of the $l$-th layer. The traditional CNN directly connects the output of the previous layer with the next layer, that is,
\begin{equation}
\bm {y}_l=F_l (\bm {y}_{l-1}).
\end{equation}
However, in DenseNet, the input of the current layer is the feature maps of all previous layers, i.e.,
\begin{equation}
  \bm {y}_l = {F_l}\left( {\left[ {{\bm {y}_0},{\bm {y}_1},...,{\bm {y}_{l - 1}}} \right]} \right),
 \end{equation}
where ${\left[ {{\bm {y}_0},{\bm {y}_1},...,{\bm {y}_{l - 1}}} \right]}$ is the concatenation of the feature maps of all these layers. The function $F_l (\cdot)$ is generally composed of multiple processing units. Similar to \cite{47}, this function in 1D-Conv-DenseNet includes three operations: batch normalization \cite{48}, rectified linear unit (ReLU) \cite{49}, and $5 \times 1$ convolution. The number of convolution kernels is set as required. The batch normalization operation is expressed as:
\begin{equation}
{\mathop{\rm BN}\nolimits} \left( x \right) = {\sigma _l}\frac{{x - {\mu _B}}}{{\sqrt {\sigma _B^2 + \varepsilon } }} + {\mu _l},
\end{equation}
where $\mu _B$ and $\sigma _B^2$ are the mean and variance calculated from the samples in the mini-batch, respectively, and $\mu _l$ and $\sigma _l$ are the offset and scale factors that are continuously updated during training. The ReLU function is defined as:
\begin{equation}
{\mathop{\rm ReLU}\nolimits} \left( x \right) = \max \left\{ {0,x} \right\} = \left\{ {\begin{array}{*{20}{c}}
{0,}\\
{x,}
\end{array}} \right.\begin{array}{*{20}{c}}
{if}\\
{if}
\end{array}\begin{array}{*{20}{c}}
{x < 0,}\\
{x \ge 0.}
\end{array}
\end{equation}
 The $P \times 1$ convolution operation can be expressed as:
\begin{equation}
{\mathop{\rm Conv}\nolimits} {\left( x \right)_{{i_{l + 1}},k}} = \sum\limits_{i = 0}^{P - 1} {\sum\limits_{c = 0}^{{C_{l}-1}} {{f_{i,c,k}} \times {x_{{i_{l + 1}} + i,c}},} }
\end{equation}
 where $f_{i,c,k}$ represents the weights of the $k$-th convolution kernel, ${\mathop{\rm Conv}\nolimits} {\left( x \right)_{{i_{l + 1}},k}}$ represents the convolution output of the corresponding position, where $0 \le {i_{l + 1}} \le {H_l} - P$, $C_l$ is the number of channels in the $l$-th layer, and $H_l$ is the length of the feature map of the convolution input. For ease of expression, we use BasicBlock($K$) to represent this basic operation of ${F_l}( \cdot )$, where $K$ represents the number of convolution kernels of the convolution layer in the module.

In CNNs, in order to maintain translation invariance and reduce the computation complexity of subsequent processing, pooling operations are often used. The size of feature maps usually becomes smaller after pooling. Only the feature maps of the same size can be depth concatenated and sent to the next layer. Therefore, DenseNet often includes two types of modules: dense module and transition module. The dense module is composed of multiple BasicBlocks. In our 1D-Conv-DenseNet, the transition module includes a batch normalization layer, a ReLU layer, a maximum pooling layer \cite{50} with filter size of $3 \times 1$ and step size of 2, and a convolution layer with kernel size of $5 \times 1$ (the number of kernels is determined as required). For simplicity, we use TransitionBlock($K$) to represent the transition module, where $K$ represents the number of convolution kernels in this module. In our designed 1D-Conv-DenseNet, first a convolutional layer containing 64 channels is used to process the input IQ signal. Then a total of four TransitionBlocks and four DenseBlocks is connected as TransitionBlock(128), DenseBlock 1, TransitionBlock(64), DenseBlock 2, TransitionBlock(64), DenseBlock 3, TransitionBlock(64), and DenseBlock 4. Each DenseBlock contains a different number of layers. Specifically, DenseBlock 1 is densely connected by two BasicBlock(128)s, DenseBlock 3 is densely connected by 4 BasicBlock(64)s, and DenseBlock 2 and DenseBlock 4 have the same structure, both of which are densely connected by three BasicBlock(64)s. After these TransitionBlocks and DenseBlocks, another convolution layer with 150 channels is used. Finally global maximum pooling and global average pooling \cite{51} are used to obtain the feature vector, according to which each binary classifier computes the probabilities of each bit being 0 or 1. Assume that the vector of the $c$-th channel sent to the global pooling layer is ${[{x_{0,c}},{x_{1,c}},...,{x_{H - 1,c}}]^T}$, then the global maximum pooling and the global average pooling operation can be expressed as
\begin{equation}
{\mathop{\rm MaxPoolGlobal}\nolimits} {(x)_c} = \mathop {\max }\limits_{0 \le i \le H - 1} {x_{i,c}},
\end{equation}
\begin{equation}
{\mathop{\rm AvgPoolGlobal}\nolimits} {(x)_c} = \frac{1}{H}\sum\limits_{i = 0}^{H - 1} {{x_{i,c}}} .
\end{equation}

There are two benefits of using global pooling. One is to maintain the maximum degree of translation invariance, that is, to adapt to the overall delay of the signal samples in time. The other benefit is the ability to adapt to changes in the length of the input signal. When the signals of different lengths are input to the network, after the operation of the global pooling layer, the output dimensions are equal to the number of input channels, that is, 150, so the dimension of the obtained feature vector is consistent (300 in this network structure). This is crucial for the DeepReceiver to adapt to multiple MCSs. After the information bit streams of the same length is coded and modulated by different MCSs, the number of symbols obtained may be different, resulting in that the number of sampling points of the received IQ signal may also be different. The network structure designed in this paper can precisely handle this situation, which also provides the possibility of using the DeepReceiver model as a unified information recovery model for multiple different MCSs.

\subsection{The Training and Inference Algorithms}
The purpose of CNN training is to optimize network parameters based on the training data set to achieve good performance on the training set, and at the same time try to make it generalizable to other data than the training set. The construction of the training set is the first step of network training. The training set for the DeepReceiver is
\begin{equation}
{\mathcal {D}} = \left\{ {\left( {{{\left[ {{\mathop{\rm Re}\nolimits}\bm{\left( r \right)},{\mathop{\rm Im}\nolimits} \bm {\left( r \right)}} \right]}^{\left( i \right)}},{\bm {s}^{\left( i \right)}}} \right)} \right\}_{i = 1}^{\mathcal {N}},
\end{equation}
where $\cal {N}$ is the number of samples in the training set. It should be pointed out that in order to make the DeepReceiver model adapt to multiple MCSs, the training set needs to include received signal samples generated with these MCSs.

The loss function is the key to training. For classification tasks, the most used loss function is cross entropy. As shown in Fig. 3, the output of our designed DeepReceiver contains $M$ binary classifiers. For simplicity, we design the loss function using the sum of the cross-entropy of the $M$ classifiers. For a minibatch containing $N_B$ samples, the loss function is defined as
\begin{equation}
 {\cal {L}} =  - \frac{1}{N_B}\sum\limits_{i = 1}^{N_B} {\sum\limits_{m = 1}^M {\sum\limits_{k = 1}^2 {{d_{imk}}\log \left( {{p_{imk}}} \right)} } } ,
\end{equation}
where ${p_{imk}}$ is the output probability of the $m$-th classifier on the $k$-th category when the $i$-th sample is used as input, and  ${d_{imk}}$ is the $k$-th true label corresponding to the $m$-th bit of the $i$-th sample. One-hot coding is used for labeling, i.e., when a  bit in the real information bit stream is 0, the corresponding label is ${\left[ {1, 0} \right]^T}$ otherwise the label is ${\left[ {0, 1} \right]^T}$. The most commonly used optimization method for deep neural network training is the stochastic gradient descent (SGD) method \cite{52}. The SGD algorithm might oscillate along the path of steepest descent towards the optimum. We use SGD with momentum which adds a momentum term to the update to reduce this oscillation \cite{53} in this paper. The update is
\begin{equation}
 \bm{w}_{t + 1} = \bm{w}_l - \tau \nabla {\cal {L}} \left( {\bm{w}_t} \right) + \vartheta \left( {\bm{w}_t - \bm{w}_{t - 1}} \right),
\end{equation}
where $\bm{w}$ is the network parameter vector (including weights and biases), $t$ is the iteration index, $\tau  > 0$ is the learning rate, and $\vartheta $ is the momentum factor, which represents the contribution of the previous gradient step to the current iteration. The gradient ${\mathcal {L}}\left( {\bm{w}_t} \right)$ is calculated and used to update the parameters using a subset of the training set, i.e., a mini-batch. In summary, the training and inference algorithms for DeepReceiver are shown in Algorithm 1.

\begin{algorithm}[htb]
\caption{Training and inference algorithms for the DeepReceiver model}
\textbf{Training procedure}\\
\hspace*{0.02in} {\bf Input:}
Training set ${\mathcal {D}} = \left\{ {\left( {{{\left[ {{\mathop{\rm Re}\nolimits}\bm{\left( r \right)},{\mathop{\rm Im}\nolimits} \bm {\left( r \right)}} \right]}^{\left( i \right)}},{\bm {s}^{\left( i \right)}}} \right)} \right\}_{i = 1}^{\mathcal {N}}$, minibatch size $N_{B}$, maximum iterations $t_{max}$, learning rate $\tau$, momentum factor $\vartheta $.
  \begin{algorithmic}
  \State Randomly initialize parameters of the network;
  \For{$t=1,2,...,t_{max}$}
¡¡¡¡\State Randomly choose $N_{B}$ samples from $\cal D$;
¡¡  \State Compute loss according to (15);
    \State Update parameters of the network according to (16);
  \EndFor\\
  \textbf{end for}
  \end{algorithmic}
\hspace*{0.02in} {\bf Output:}
DeepReceiver model ${\cal {F}}(\cdot;\cal {Q})$.\\
\textbf{Inference procedure}\\
\hspace*{0.02in} {\bf Input:}
 The newly received IQ signal waveform, the trained DeepReceiver model ${\cal {F}}(\cdot;\cal {Q})$.
\begin{algorithmic}
 \State Compute the SoftMax output of each binary classifier $p_{m0}$ and $p_{m1}$;
 \For{$m=1,2,..., M$}
 \If {$p_{m0} > p_{m1}$}
    \State ${\widehat s_m} = 0$;
 \Else
    \State  ${\widehat s_m} = 1$.
 \EndIf
 \EndFor\\
  \textbf{end for}
\end{algorithmic}
 \hspace*{0.02in} {\bf Output:}
 Information bit stream $\bm{\widehat s} = {[{\widehat s_1},{\widehat s_2},...,{\widehat s_M}]^T}.$
\end{algorithm}

\subsection{Complexity Analysis}
Since training can be performed offline, our main concern is the inference complexity after the model is deployed. In 1D-Conv-DenseNet, the computational complexity of each convolutional layer is
\begin{equation}
{{\cal C}_{{\rm{Conv}}}} \sim O\left( {{H_l}{C_l}{P_l}{K_l}} \right),
\end{equation}
where ${H_l} \times {C_l}$ represents the size of the input feature map, $P_l$ represents the size of the convolution kernel, and $K_l$ represents the number of convolution kernels. The computational complexity of the batch normalization layer and the ReLU layer are both
\begin{equation}
{{\cal C}_{{\rm{ReLU}}}} \sim O(H_l C_l).
\end{equation}
The computational complexity of the pooling layer is
\begin{equation}
{{\cal C}_{{\rm{Pooling}}}} \sim O\left( {{H_l}{C_l}{F_l}/{D_l}} \right),
\end{equation}
where $F_l$ is the size of the pooling filter and $D_l$ is the downsampling factor. In 1D-Conv-DenseNet, the maximum number of convolution kernels is limited. When the input signal length is $N$, the overall computational complexity of the network is
\begin{equation}
{{\cal C}_{{\cal {F}}(\cdot;\cal {Q})}} \sim O(N).
\end{equation}

The storage complexity includes two aspects, one is the storage of network parameters, and the other is the storage of the feature maps during the inference process. Since the feature map of the previous layer can be overwritten after the calculation is completed, the storage capacity of the feature maps is twice the largest feature map. When the input signal length is $N$ and the number of recovered information bits is $M$, the number of the network parameters of 1D-Conv-DenseNet is
\begin{equation}
\left| {\cal Q} \right| = 1248322+602M,
\end{equation}
and the maximum feature map size is
\begin{equation}
\mathop {\max }\limits_l \left\{ {\left| {\bm{y}_{l}} \right|} \right\} = \frac{N}{2} \times 384 = 192N.
\end{equation}
Therefore, the total parameter storage is
\begin{equation}
{{\cal S}_{{\cal {F}}(\cdot;\cal {Q})}} = \left| {\cal Q} \right|+2 \times \mathop {\max }\limits_l \left\{ {\left| {\bm{y}_{l}} \right|} \right\}=1248322+602M+384N.
\end{equation}

For time-sensitive applications, further consideration can be given to adopting methods such as network pruning to further reduce the computational complexity and storage complexity of the neural network model. For example, layer-wise pruning \cite{60} can be performed to reduce the number of neural network layers, thereby reducing inference delay.

\section{Performance Analysis}
In this section, we demonstrate the performance of the DeepReceiver through simulation experiments. The simulation setting is firstly described. Then, the simulation results of DeepReceiver with the effects of noise (AWGN and AGGN), RF impairments (frequency deviation and IQ imbalance), channel fading (frequency flat Rayleigh fading and frequency selective Rayleigh fading), cochannel interference (single-tone, MSK and BPSK), dynamic environment and unified blind reception are discussed.
\subsection{ Simulation Setting}
Two modulations are considered in the simulation: BPSK and quadrature phase shift keying (QPSK). The channel coding uses (7,4) Hamming coding, the pulse shaping uses a raised cosine filter with roll-off factor 0.5. For the BPSK + Hamming method, the number of bits in the information bit stream is 32. The information bit stream is randomly generated, channel encoded into 56 bits, and then sent out after being mapped by BPSK modulation and shaped by raised cosine filtering. The sampling rate of the received signal is 8 times the symbol rate, that is, the number of sampling points per symbol is 8. The timing of the first sampling point is randomly selected within 1/8 times the symbol period. The data length of each IQ sample obtained is thus 448. With QPSK modulation, (7,4) Hamming code and raised cosine filter are also used. The number of bits in the original information bit stream is 32, the timing of the first sampling point is randomly selected within 1/8 times the symbol period, and the oversampling ratio of the received signal is also 8, so the data length of each IQ sample obtained is 224. Note that this is the configuration used in most simulations in this paper. For fading channel and dynamic environment experiments, the signal length used will be different. For multiple MCSs experiment, the modulation and coding will be different. We will give the details in the corresponding experiments.

Both the training and test data sets are generated with simulation. The non-ideal conditions considered in each experiment are different. For each experiment, the corresponding non-ideal conditions will be simulated in generating the IQ data. In each training data set, the signal $E_b/N_0$ ranges from 0 dB to 8 dB, with an interval of 1 dB. The number of data samples per $E_b/N_0$ is 200,000, so the total sample size is 1,800,000. In each test data set, the signal $E_b/N_0$ ranges from 0 dB to 8 dB with an interval of 0.5 dB, and the number of samples per $E_b/N_0$ is 200,000. The reason why more $E_b/N_0$ is selected than the training set is to test the adaptability of the model to the untrained $E_b/N_0$. Training is performed on a Nvidia V100 GPU. All parameters of the 1D-Conv-DenseNet are randomly initialized with a Gaussian distribution. The SGD method with momentum is used for training, and the momentum factor is 0.9. During training, the mini-batch size is 256, the number of epochs is 8, and the initial learning rate is 0.001. After every 2 epochs, the learning rate is reduced to 1/10 of the previous learning rate.

\subsection{Effects of Noise: AWGN and AGGN}
\begin{figure}[!t]
\centering
\includegraphics[width=3.4in]{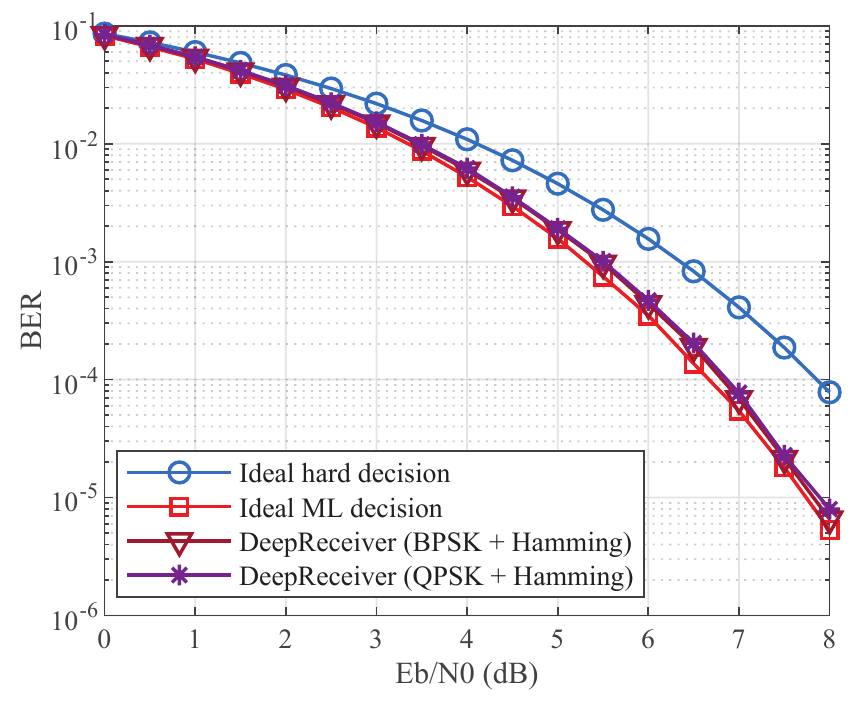}
\caption{BER performance in the case of AWGN. BPSK + Hamming and QPSK + Hamming are used for modulation and channel coding.}
\label{Fig.3}
\end{figure}
\emph{1) AWGN:}  We first consider the performance under the AWGN. Consider BPSK + Hamming and QPSK + Hamming. The DeepReceiver models are trained separately for the two modulations. Fig. 3 illustrates the simulation results. For comparison, the results of ideal hard decision and ideal maximum likelihood (ML) decision are also shown in the figure. Among them, the ideal hard decision refers to the method of demodulating under perfect assumptions and then inputting the demodulated bit stream to a Hamming decoder with hard decision without the influence of any other factors except AWGN. Ideal ML decision refers to the ML decoding method with ideal assumptions, i.e., in the absence of any other non-ideal factors except AWGN. Since the simulated information bit stream follows an equal probability distribution, the ideal ML decision also represents the best performance we can obtain under ideal conditions. Unless otherwise specified, when referring to both the ideal hard decision and the ideal ML decision hereafter, we refer to the methods under these ideal assumptions. It should be noted in advance that all the hard decision methods in the subsequent results of this paper assume that the symbol timing is ideal, that is, the problem of symbol synchronization need not be considered. However, the DeepReceiver need to automatically learn and implement symbol synchronization from IQ sequences with timing errors. It can be seen from Fig. 3 that for both BPSK + Hamming and QPSK + Hamming, the performance of the DeepReceiver is very close to the ideal ML decision and far better than the traditional hard decision method, which illustrates the DeepReceiver¡¯s potential to approach performance limits. On the untrained $E_b/N_0$s, the DeepReceiver also achieves performance close to ML decision, indicating that it has a good generalization ability for $E_b/N_0$s. It should be noted that, for the BPSK + Hamming and QPSK + Hamming, the BER performance of the ideal hard decision is the same, and the BER performance of the ideal ML decision is also the same, so it is not specifically indicated in the figure whether it is BPSK or QPSK.
\begin{figure}[!t]
\centering
\includegraphics[width=3.4in]{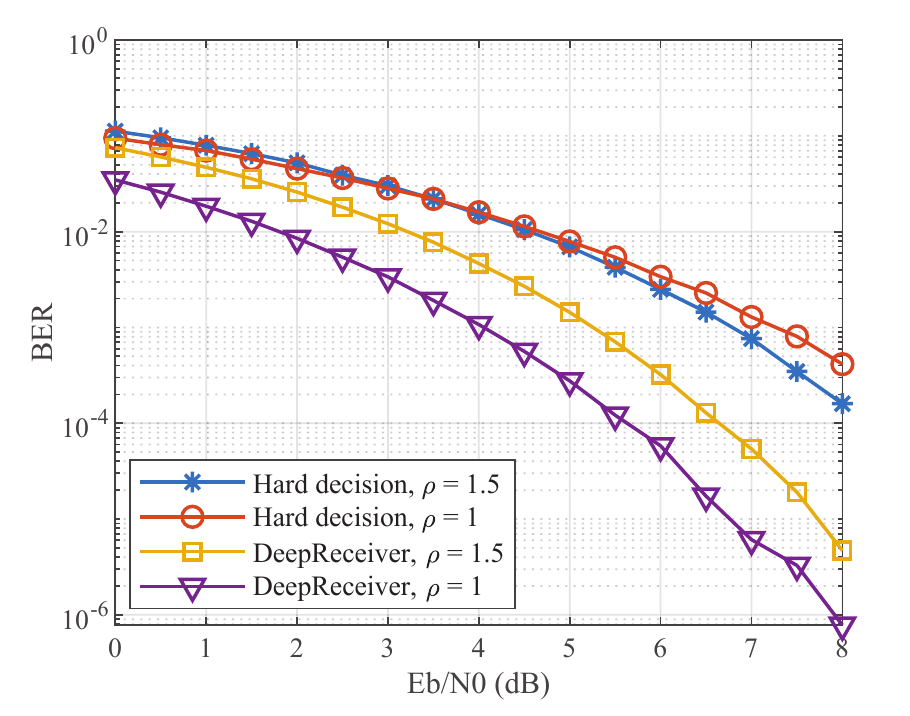}
\caption{ BER performance in the case of AGGN.}
\label{Fig.4}
\end{figure}
\begin{figure}[!t]
\centering
\includegraphics[width=3.4in]{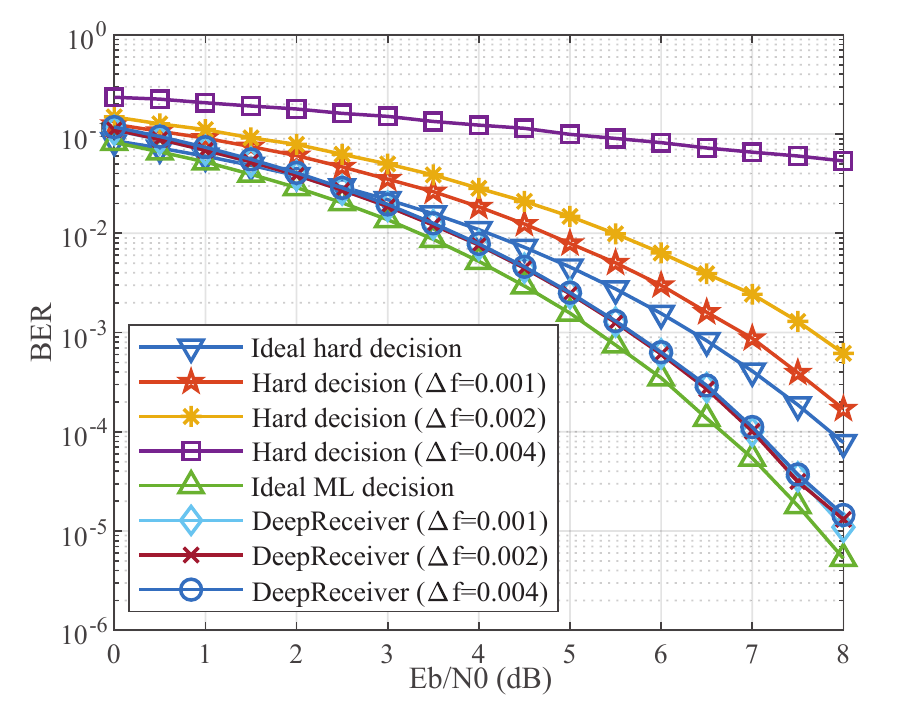}
\caption{Performance under different carrier frequency deviations.}
\label{Fig.5}
\end{figure}

\emph{2) AGGN:} The above considers the situation of AWGN, but in the actual wireless communication process, the noise does not necessarily conform to the characteristics of AWGN. In order to evaluate the DeepReceiver with other noise distribution, we performed a simulation experiment on the performance of the algorithm under AGGN. The AGGN parameters are set as follows: $\mu = 0$, $\gamma = 1$, and $\rho = 1.5$ in case 1 and $\rho = 1$ in case 2. Fig. 4 shows the simulation results. It can be seen that under AGGN, the performance of the traditional hard decision method decreases, and the performance at $\rho = 1$ is worse than that at $\rho  = 1.5$. This is because the hard decision method assumes AWGN distribution and the distribution of the AGGN is farther away from AWGN at $\rho = 1$ than $\rho = 1.5$. The performance of the DeepReceiver is better than the traditional method under both parameter settings. What is interesting is that, unlike the traditional hard decision method, the performance of the DeepReceiver is better when $\rho = 1.5$ than when $\rho = 1$, which indicates that the DeepReceiver has learned a receiving method that better matches the noise distribution.

\subsection{ Effects of RF Impairments: Frequency Deviation and IQ Imbalance}
\emph{1) Frequency deviation:} In wireless communication system, two independent local oscillators are used at the transmitter and at the receiver. There may be a certain deviation in their frequencies. In addition, when there is relative movement between the transmitter and the receiver, a Doppler shift will occur. Under the influence of these factors, there will be a certain frequency deviation between the received signal and the transmitted signal. We analyze the performance of the DeepReceiver in the presence of carrier frequency deviation. In the simulation, the normalized carrier frequency offset $\Delta f$ (relative to the symbol rate) is randomly generated within the range of [-0.01, 0.01]. The other settings are the same as those in the previous simulation. Fig. 5 illustrates the results on the test set. The performance of the traditional hard decision method is used for comparison. It can be seen that the traditional hard decision is greatly affected by the carrier frequency deviation. As the carrier frequency deviation increases, the performance deteriorates significantly, especially when $\Delta f = 0.004$, the BER is worse than 0.01 when $E_b/N_0$ is in the range of 0-8 dB. Such BER performance will be difficult to meet the needs of practical applications. However, the BER performance of the DeepRceiver is still very close to the ideal ML decision, indicating that it can overcome the influence of carrier frequency deviation to a certain degree.
\begin{figure}[!t]
\centering
\includegraphics[width=3.4in]{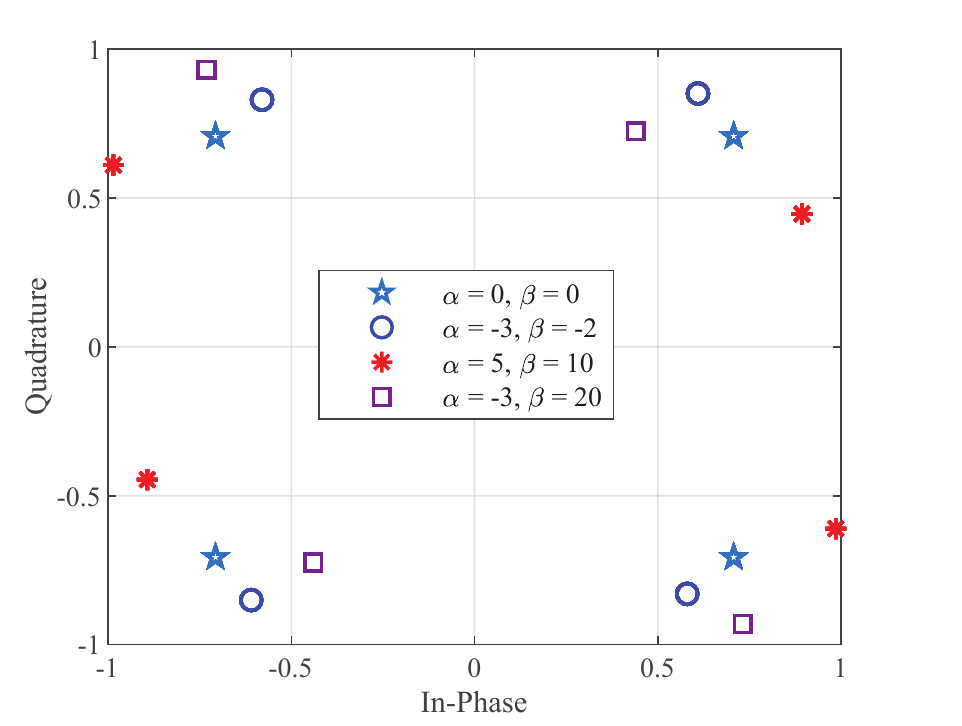}
\caption{Constellation of QPSK in the ideal case and the constellations of QPSK with IQ imbalance. In the figure, $\alpha$ is amplitude imbalance in dB and $\beta$ is phase imbalance in degrees.}
\label{Fig.6}
\end{figure}

\emph{2) IQ imbalance:} Due to the non-ideality of RF devices, the received IQ signal may have an IQ imbalance, that is, an imbalance in the amplitude and/or phase of the I channel and the Q channel. IQ imbalance can be described by a set of parameters ($\alpha$, $\beta$), where $\alpha$ is amplitude imbalance in dB and $\beta$ is phase imbalance in degrees. We analyze the performance of the DeepReceiver in the presence of IQ imbalance. The simulation uses QPSK + Hamming parameter settings and considers three IQ imbalance configurations, (-3,-2), (5,10), and (-3,20), respectively. Fig. 6 shows the constellation of QPSK in the ideal case and the constellations of QPSK in the with IQ imbalances. It can be seen that the IQ imbalance causes distortion to the ideal QPSK constellation.
\begin{figure}[!t]
\centering
\includegraphics[width=3.4in]{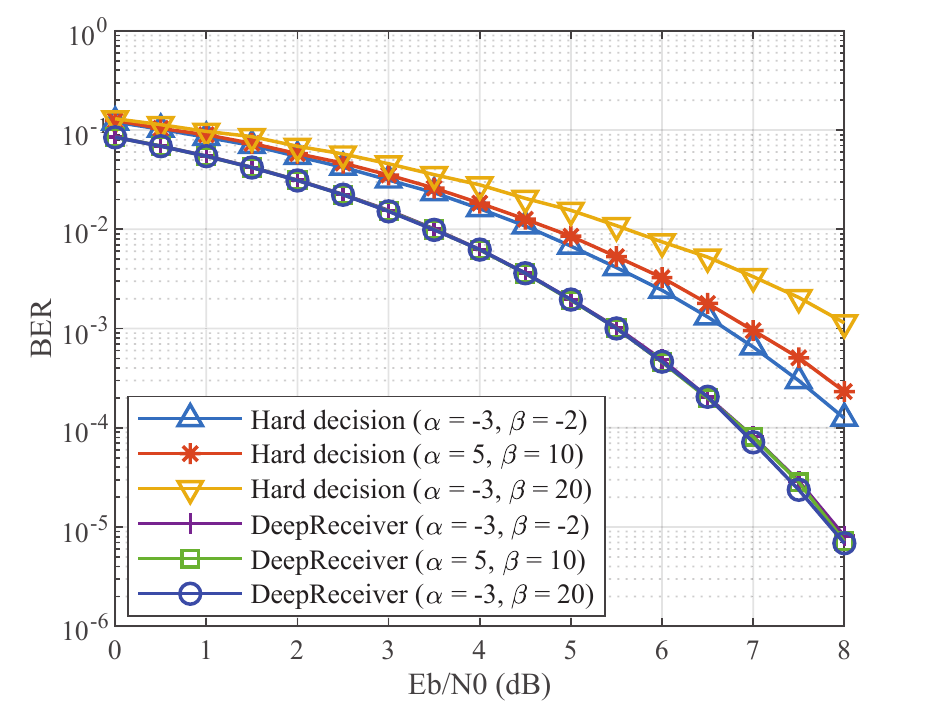}
\caption{ Performance under IQ imbalance. In the figure, $\alpha$ is amplitude imbalance in dB and $\beta$ is phase imbalance in degrees.}
\label{Fig.7}
\end{figure}

Fig. 7 shows the experimental results. It can be seen that for the traditional hard decision method, as the IQ imbalance increases, the BER performance becomes worse. However, in all three cases of IQ imbalance, the performance obtained by the DeepReceiver is very close to that without the IQ imbalance, which indicates that the DeepReceiver can learn to correct the impact of the IQ imbalance automatically.

\subsection{Effects of Channel Fading: Frequency Flat Fading and Frequency Selective Fading}
During the transmission of communication signals, due to factors such as geographical environment and obstacles, the signals received by the receiver may be influenced by multipath fading. We perform simulation analysis on the performance of the DeepReceiver in multipath fading channels. The simulations assume a symbol rate of 1Msps. Two Rayleigh fading channels are considered: frequency flat Rayleigh fading channel and frequency selective Rayleigh fading channel. Traditional equalization methods usually require a known training sequence of a certain length, which is added in front of the information payload. The simulation uses BPSK modulation and Hamming coding. The number of original information bits is 32, which is randomly generated, and channel encoded to 56 bits. What is different from the previous simulations is that in order to compare the performance with the traditional equalization methods, we add a fixed $L$-bit sequence in front of the channel encoded bit stream. The oversampling ratio of the received IQ signal is also 8 and the length of a signal sample is thus 448 + 8$L$. DeepReceivers are trained separately for the two Rayleigh fading channels.
\begin{figure*}[!t]
\subfigure[]{
\includegraphics[width=3.4in]{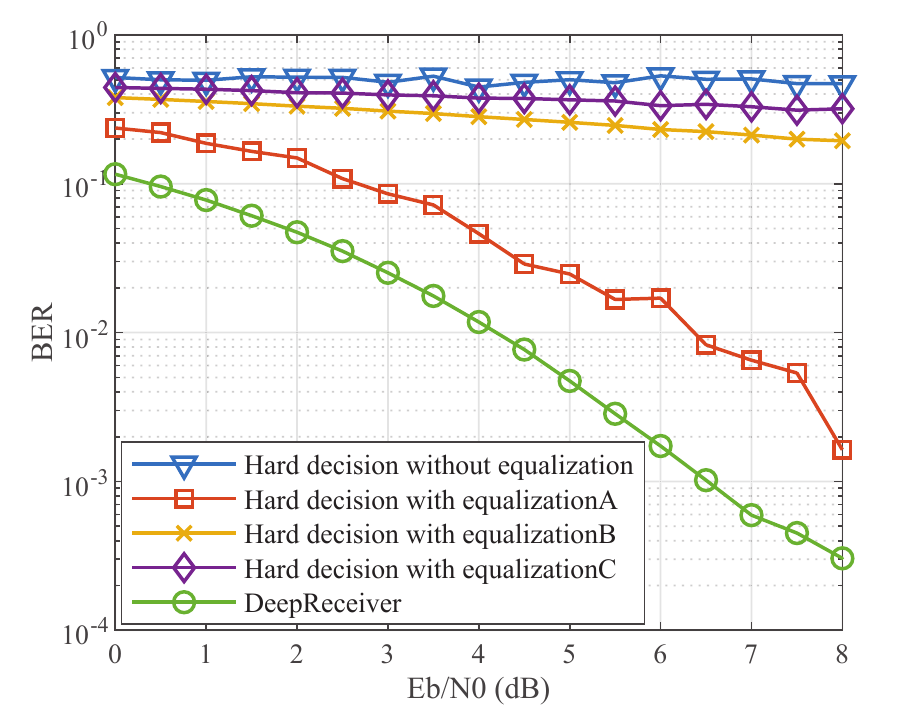}
}
\subfigure[]{
\includegraphics[width=3.4in]{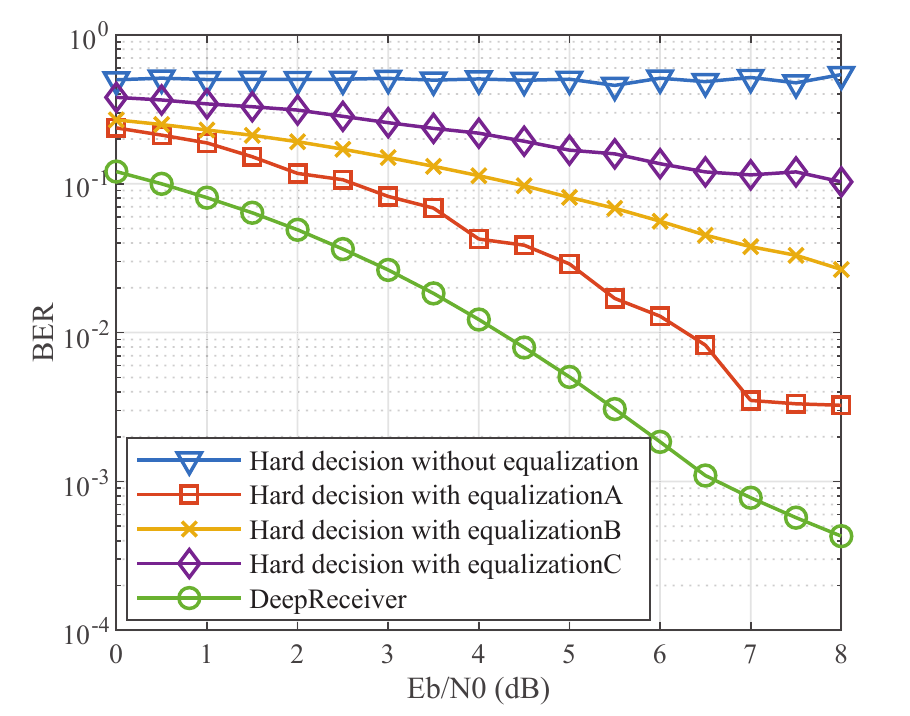}
}
\caption{Performance under frequency flat Rayleigh fading channel. (a) $L$ = 8 and (b) $L$ = 32.}
\label{Fig.8}
\end{figure*}
\begin{figure*}[!t]
\subfigure[]{
\includegraphics[width=3.4in]{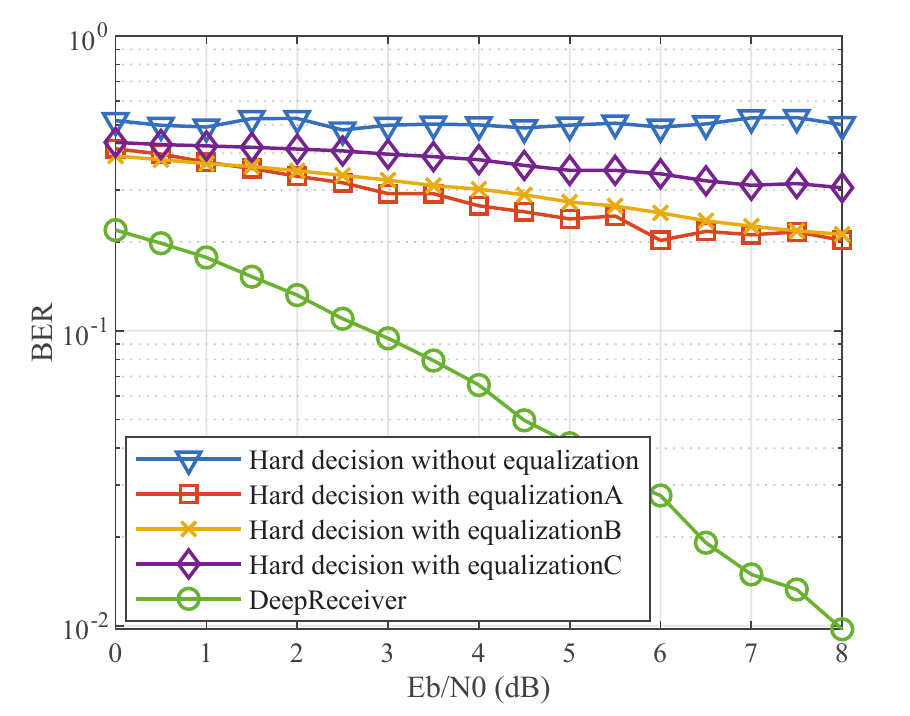}
}
\subfigure[]{
\includegraphics[width=3.4in]{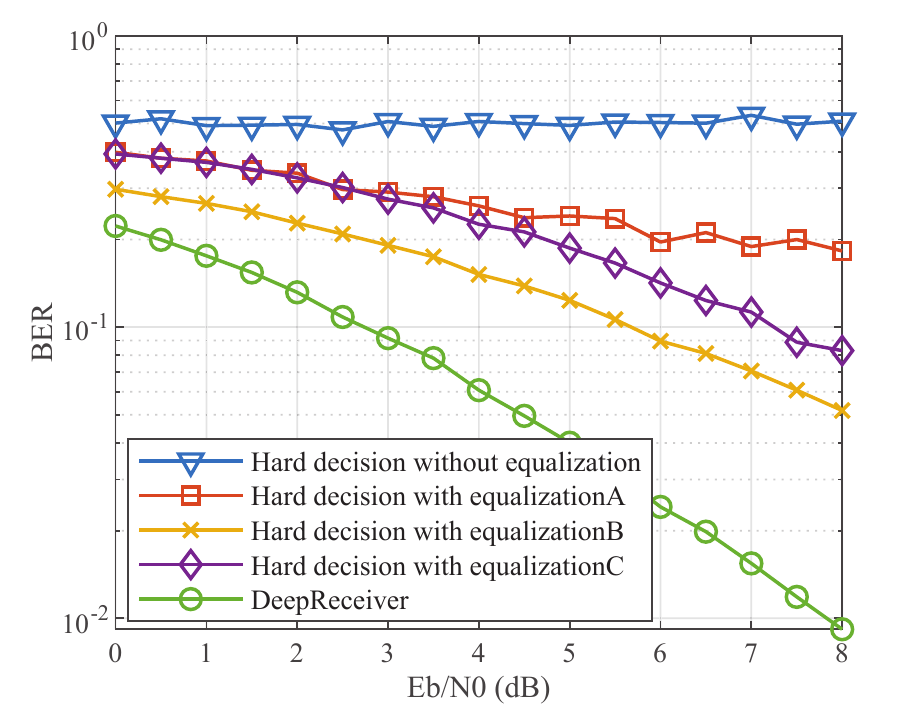}
}
\caption{Performance under frequency selective Rayleigh fading channel. (a) $L$ = 8 and (b) $L$ = 32.}
\label{Fig.9}
\end{figure*}

As a method for performance comparison, adaptive equalization + BPSK demodulation + Hamming hard decision decoding is adopted. Among them we consider three adaptive equalization algorithms. The first equalization algorithm, denoted as equalizationA, is a linear equalization. The least mean square (LMS) algorithm is used. The number of taps is 1 and the step size for LMS is 0.01. The second equalization algorithm, denoted as equalizationB, is also a linear equalization, using a recursive least squares (RLS) algorithm with 8 taps. The reference tap is 3 and the RLS forgetting factor is 0.99. The third equalization algorithm, denoted as equalizationC, is a decision feedback equalization algorithm, using a 6-tap forward filter and a 2-tap inverse filter, with a reference tap of 3. It also uses RLS with a forgetting factor of 0.99.

\emph{1) Frequency flat Rayleigh fading:} For frequency-flat Rayleigh fading, the maximum Doppler shift is set to 30 Hz.
Fig. 8 shows the simulation results for the frequency flat Rayleigh fading channel. It can be seen that among the three traditional adaptive equalization algorithms, equalizationA performs the best. The comparison of the results under different $L$ shows that under the two training sequence lengths, equalizationA performs similarly. The performances of hard decision methods using equalizationB and equalizationC improve with the increase of $L$, but are still much worse than the hard decision method using equalizationA. Under this fading channel, the performance of the DeepReceiver under the two $L$ settings is similar. Its performance is far better than the three hard decision methods with adaptive equalization, which shows its superiority under flat Rayleigh fading channels.
\begin{figure}[!t]
\begin{center}
\subfigure[]{
\includegraphics[width=0.22\textwidth]{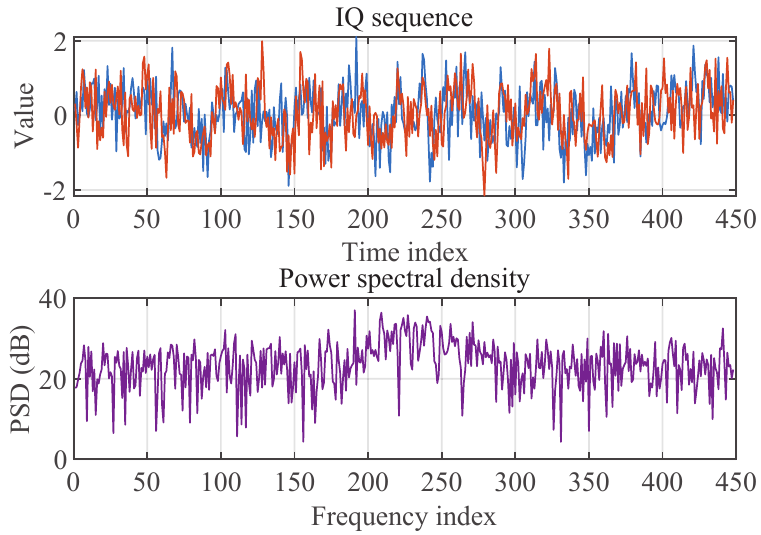}
}
\subfigure[]{
\includegraphics[width=0.22\textwidth]{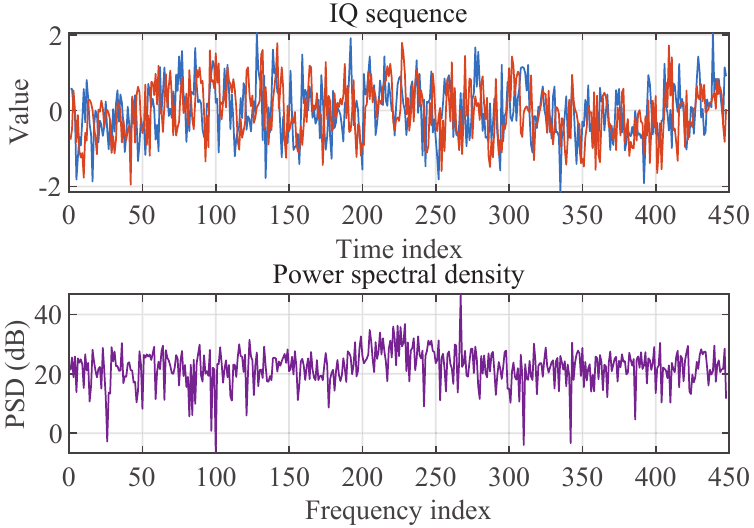}
}
\subfigure[]{
\includegraphics[width=0.22\textwidth]{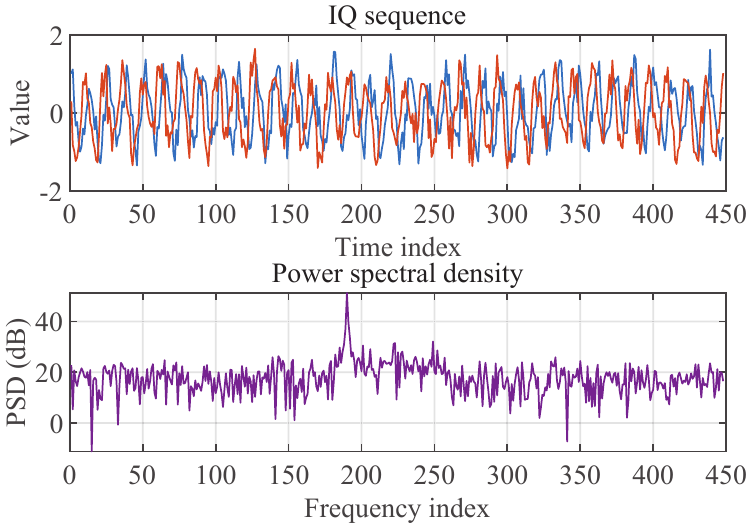}
}
\subfigure[]{
\includegraphics[width=0.22\textwidth]{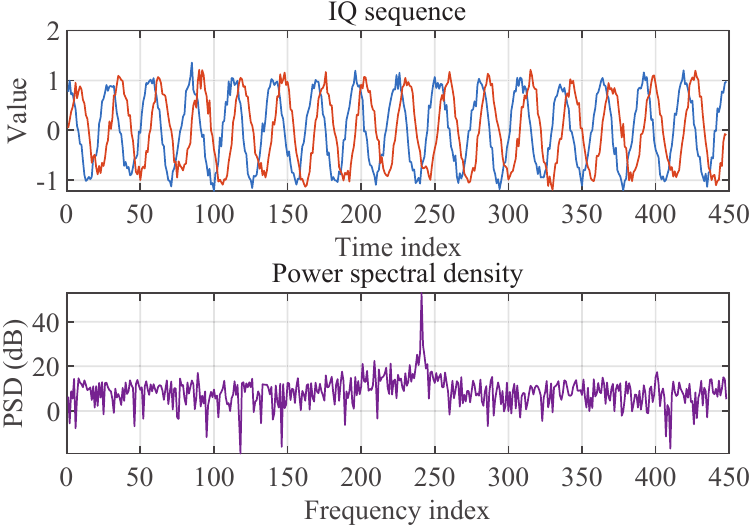}
}
\end{center}
\caption{Time-domain and frequency-domain plots of some samples, $E_b/N_0$ = 8 dB. (a) ISR = -10 dB, (b) ISR = 0 dB, (c) ISR = 10 dB, and (d) ISR = 20
dB.}
\label{Fig.10}
\end{figure}

\emph{2) Frequency selective Rayleigh fading:} In frequency selective Rayleigh fading, The maximum Doppler shift is 30 Hz, the number of paths is 3, the path delays are 0 seconds, 90 microseconds, and 1.5 microseconds, respectively, and the average path gain is 0 dB, -3 dB, and -6 dB, respectively.
Fig. 9 shows the simulation results of the frequency selective Rayleigh fading channel. When $L = 8$, the performance of traditional hard decision methods with adaptive equalization is very poor. When $L$ increases to 32, the hard decision with equalizationB shows a certain performance advantage among the three adaptive equalization algorithms. When $L = 32$, the performance of the DeepReceiver is slightly improved compared with $L = 8$. Under both $L$ settings, the performance of the DeepReceiver is far better than these three hard decision methods with adaptive equalizations, and the performance gain is more obvious than that of the flat fading channel, which validates the DeepReceiver¡¯s superiority in frequency selective Rayleigh fading channel.

\subsection{Effects of Cochannel Interference: Single-tone Interference, MSK Interference and BPSK Interference}
During the communication process, the communication system may be unintentionally or intentionally interfered by other emitters. Without corresponding anti-interference measures, the performance of the communication system will deteriorate seriously when being interfered. In general, the larger the interference power, the more severe the performance degradation. In this section we analyze the performance of the DeepReceiver in the presence of co-channel interference. We use BPSK + Hamming to generate the communication signal and consider three types of interference: single-tone interference, MSK interference and BPSK interference.

\emph{1) Single-tone interference:} We first consider cochannel single-tone interference in the simulation. The frequency of single-tone interference is randomly generated within the signal bandwidth, and the power of single-tone interference is generated according to the interference-to-signal power ratio (ISR) which is within the range of [-20 dB, 30 dB]. The signal $E_b/N_0$ ranges from 0 to 8 dB with an interval of 1 dB. Fig. 10 shows the time-domain IQ sequence and frequency-domain power spectral density (PSD) of some signal samples. When the ISR is high, the single-tone interference is clearly visible on the PSD diagram.
\begin{figure}[!t]
\subfigure[]{
\includegraphics[width=3.4in]{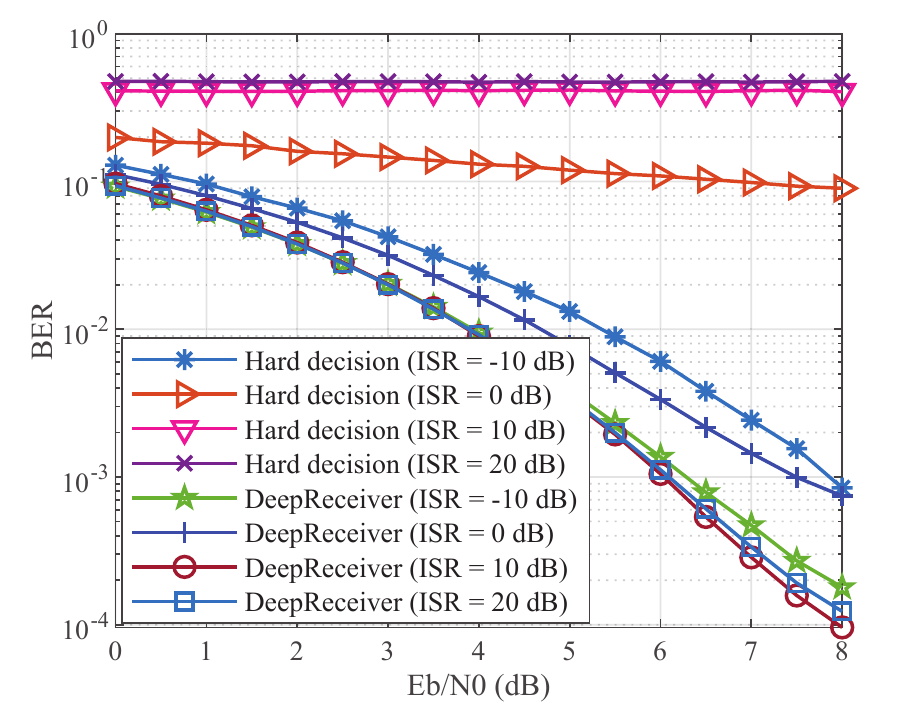}
}
\subfigure[]{
\includegraphics[width=3.4in]{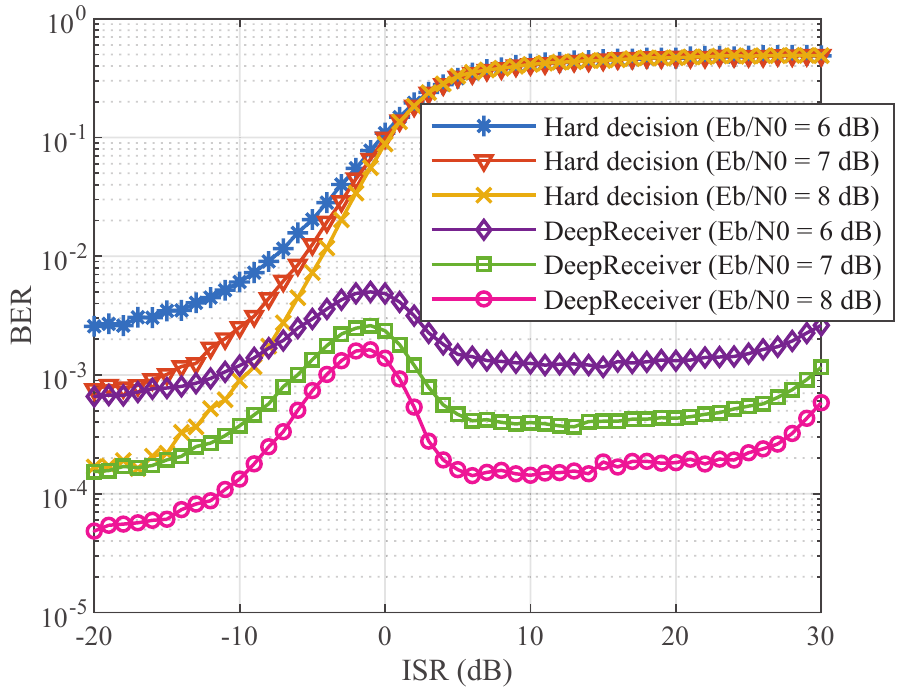}
}
\caption{Performance under cochannel single-tone interference. (a) Curves of the BER with $E_b/N_0$ under specific ISRs and (b) curves of the BER with ISR under specific $E_b/N_0$s.}
\label{Fig.11}
\end{figure}

Fig. 11 (a) and Fig. 11 (b) show the curves of the BER with $E_b/N_0$ under specific ISRs, and the curves of the BER with ISR under specific $E_b/N_0$s, respectively. It can be seen from the figure that when the ISR increases, the performance of the traditional hard decision method gradually decreases. When ISR $> 0$ dB, the BER $> 0.1$, and when the ISR $> 10$ dB, the BER approaches 0.5, and the algorithm almost completely failed. For the DeepReceiver, it has certain anti-interference ability under various ISRs. Especially when the ISR is in the range of 5 dB to 25 dB, the anti-interference effect of the DeepReceiver is obvious. It is interesting that for DeepReceiver, it is not that the larger the ISR, the worse the performance. When the interference power is close to the signal power, the performance of the DeepReceiver is worse than when the ISR is greater, but it is still far superior to the traditional method. A possible reason for this phenomenon is that when the ISR is large, it is easier to estimate the frequency, phase, amplitude of the single-tone interference. By subtracting the reconstructed interference signal from the received mixed signal, information can be recovered from the remaining BPSK signals. When the ISR is near 0 dB, it is difficult to estimate the parameters of the interference signal. These results verify the excellent performance of the DeepReceiver in overcoming cochannel single-tone interference.

\emph{2) MSK interference:} Similar to the previous simulation, when the training data set are generated, the center frequency of MSK is randomly generated within the BPSK signal bandwidth, the symbol rate of MSK is 8/5 times the symbol rate of BPSK, the power of the MSK interference is randomly generated within the range of [-20 dB, 30 dB] according to the ISR, and the BPSK signal $E_b/N_0$ ranges from 0 dB to 8 dB with an interval of 1 dB. The test data set is generated the similar way. The difference is the BPSK signal $E_b/N_0$ ranges from 0 dB to 8 dB with an interval of 0.5 dB. Fig. 12 shows the curves of the BER with $E_b/N_0$ under specific ISRs, and the curves of the BER with ISR under specific $E_b/N_0$s, respectively. Note that these results are obtained from the test data set. As can be seen from the figure, similar to single-tone interference, with the increase of ISR, the performance of the traditional method gradually deteriorates under MSK interference, especially when ISR $> 10$ dB, the BER deteriorates to nearly 0.5. For the DeepReceiver, it has certain anti-interference ability under these ISRs. Especially when the ISR is in the range of 10 dB to 25 dB, the DeepReceiver is less affected by the MSK interference. These results verify the excellent performance of the DeepReceiver in overcoming co-channel MSK interference.

\begin{figure*}[!ht]
\subfigure[]{
\includegraphics[width=3.4in]{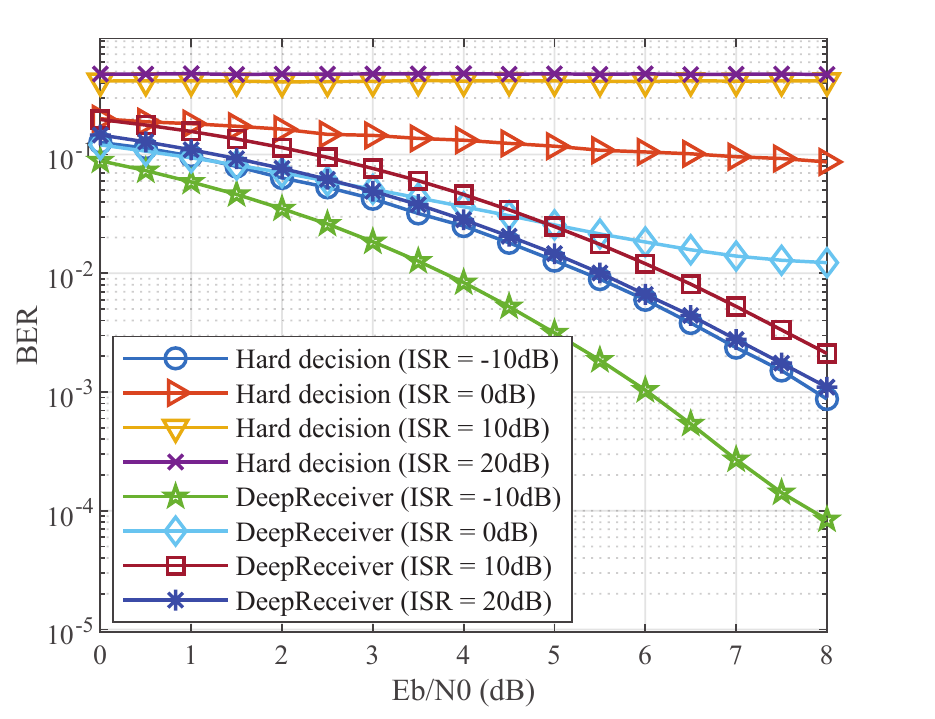}
}
\subfigure[]{
\includegraphics[width=3.4in]{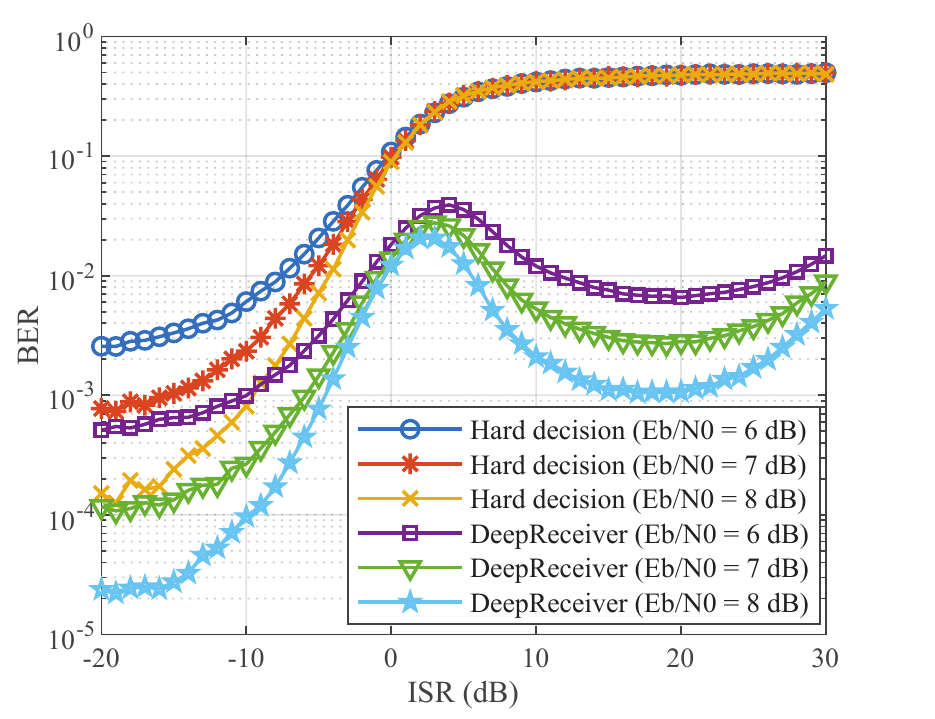}
}
\caption{Performance under cochannel MSK interference. (a) Curves of the BER with $E_b/N_0$ under specific ISRs and (b) curves of the BER with ISR under specific $E_b/N_0$s.}
\label{Fig.12}
\end{figure*}
\begin{figure*}[!ht]
\subfigure[]{
\includegraphics[width=3.4in]{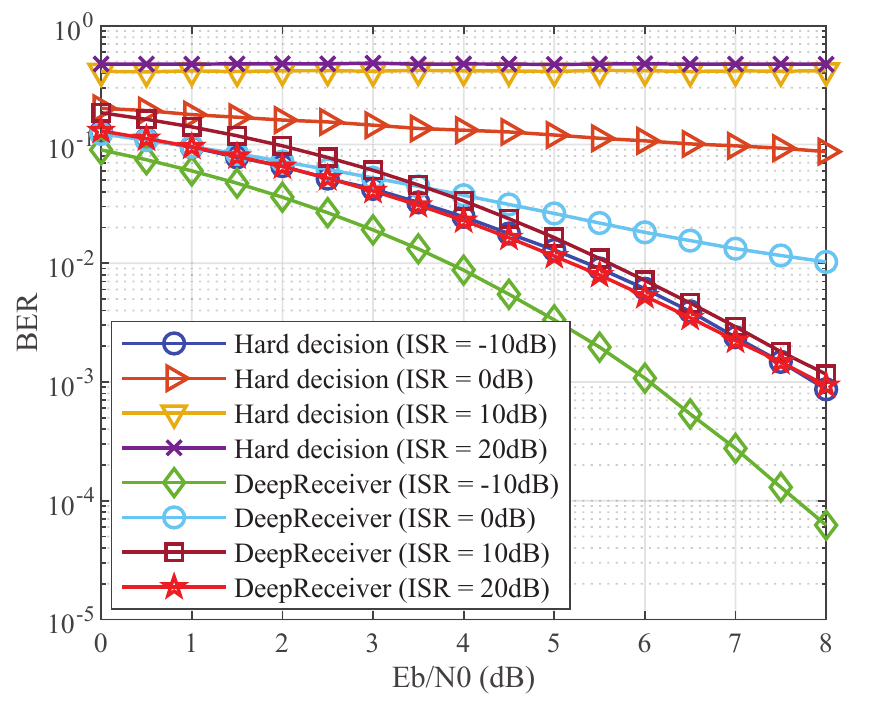}
}
\subfigure[]{
\includegraphics[width=3.4in]{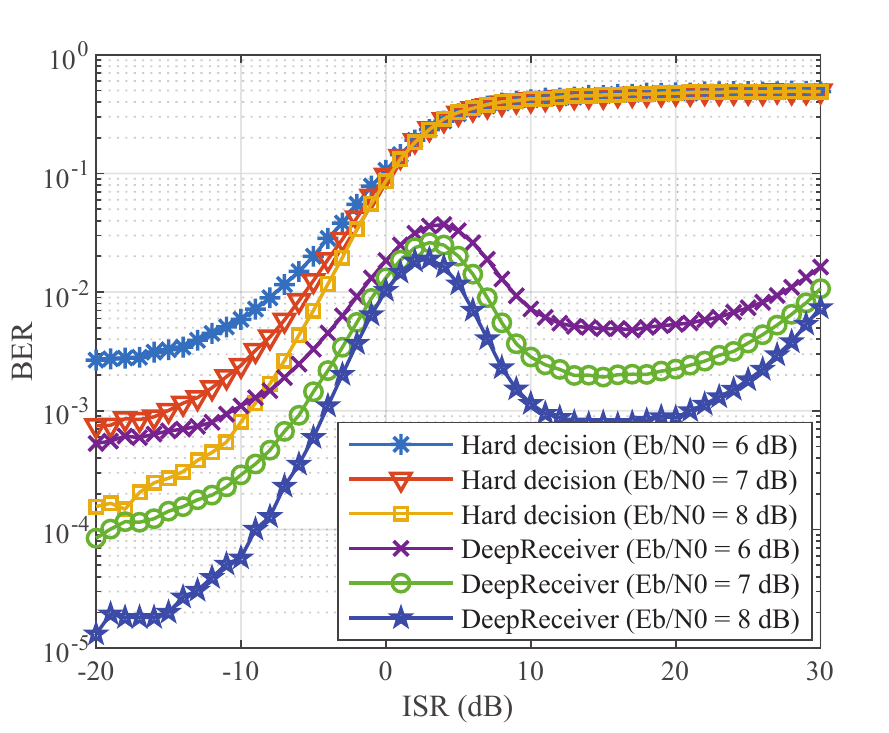}
}
\caption{Performance under cochannel BPSK interference. (a) Curves of the BER with $E_b/N_0$ under specific ISRs and (b) curves of the BER with ISR under specific $E_b/N_0$s.}
\label{Fig.13}
\end{figure*}

\emph{3) BPSK interference}: Finally, we give the anti-jamming performance of the DeepReceiver when the interference signal and the communication signal are of the same modulation, i.e., BPSK in this case. The interference power is randomly generated within the range of [-20 dB, 30 dB], the center frequency of the interference is randomly generated within the communication signal bandwidth, and the signal $E_b/N_0$ ranges from 0 dB to 8 dB. The symbol rate of the interference is 8/7 times of symbol rate of the communication signal. A raised cosine filter is used for pulse shaping of the interference and the roll-off factor is 0.3. Fig. 13 shows the curves of the BERs. It can be seen that with the increase of ISR, the performance of traditional methods gradually deteriorates. However, although the interference signal and communication signal are both BPSK signals, the DeepReceiver can still overcome the influence of interference, which further verifies the excellent performance of the DeepReceiver.

\subsection{ Performance in Dynamic Environment}
In the above experiments, the DeepReceiver model in various scenarios was trained separately. In order to verify the adaptability of the DeepReceiver to dynamic environment, we conduct unified training of DeepReceiver model in this experiment. The training set contains the signal samples under various signal-to-noise ratios in four scenarios: AWGN, AGGN, frequency flat Rayleigh fading, and frequency selective Rayleigh fading. The number of fixed bits added for equalization is $L=32$. After the training is completed, the performance of the DeepReceiver is tested on the new test set. The results show that the performance in the four scenarios is consistent with the performance obtained when the model is trained separately. For simplicity, here we will not repeat the curve of BER with $E_b/N_0$. Instead, we show the performance under a dynamic environment consists of four specific settings in Fig. 14. The four settings are:
\begin{itemize}
  \item Setting 1: $E_b/N_0$ = 6 dB, AWGN;
  \item Setting 2: $E_b/N_0$ = 6 dB, AGGN;
  \item Setting 3: $E_b/N_0$ = 6 dB, frequency flat Rayleigh fading;
  \item Setting 4: $E_b/N_0$ = 7 dB, frequency selective Rayleigh fading.
\end{itemize}

In this dynamic environment, the channel changes with time from Setting 1 to Setting 2, then to Setting 3, and finally to Setting 4. In each scenario, the number of data blocks sent is 100,000, and the BER is calculated using 10,000 data blocks. It can be seen from the results that the performance of the DeepReceiver is better than the other four traditional methods in all four settings. Traditional methods have different performances in four scenarios. For example, in Setting 1 and Setting 2, the hard decision without equalization method performs best, but in Setting 3 and Setting 4, the performance of this method deteriorates significantly, which is worse than the other three methods with equalization. Comparing the three equalization methods, we can see that in Settings 1, 2 and 3, equalizationA performs best, but in setting 4, equalizationA performs worse than equalizationB and equalizationC. Therefore, it is difficult to choose a specific traditional method that performs best in all four Settings. However, the DeepReceiver can obtain good performance under four settings without knowing the specific setting it is currently in. It has a good adaptability to the dynamic environment.
\subsection{Unified Blind Reception for Multiple MCSs}
\begin{figure}[!t]
\includegraphics[width=3.4in]{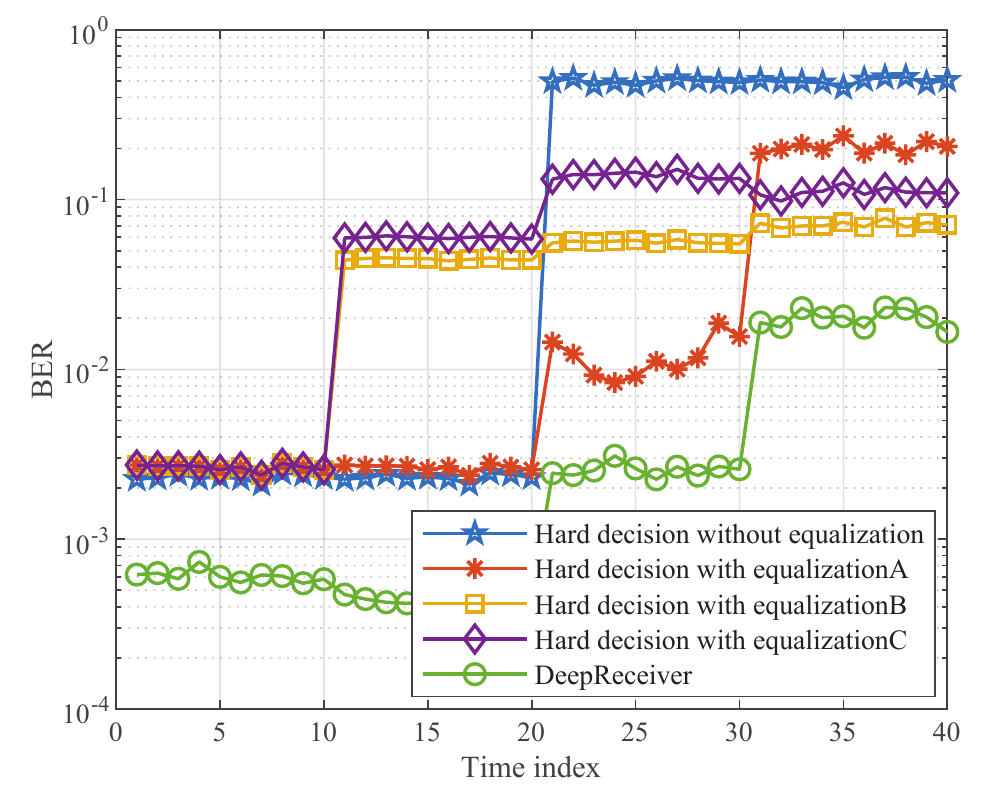}
\caption{Performance in dynamic environment.}
\label{Fig.14}
\end{figure}
\begin{figure}[!t]
\includegraphics[width=3.4in]{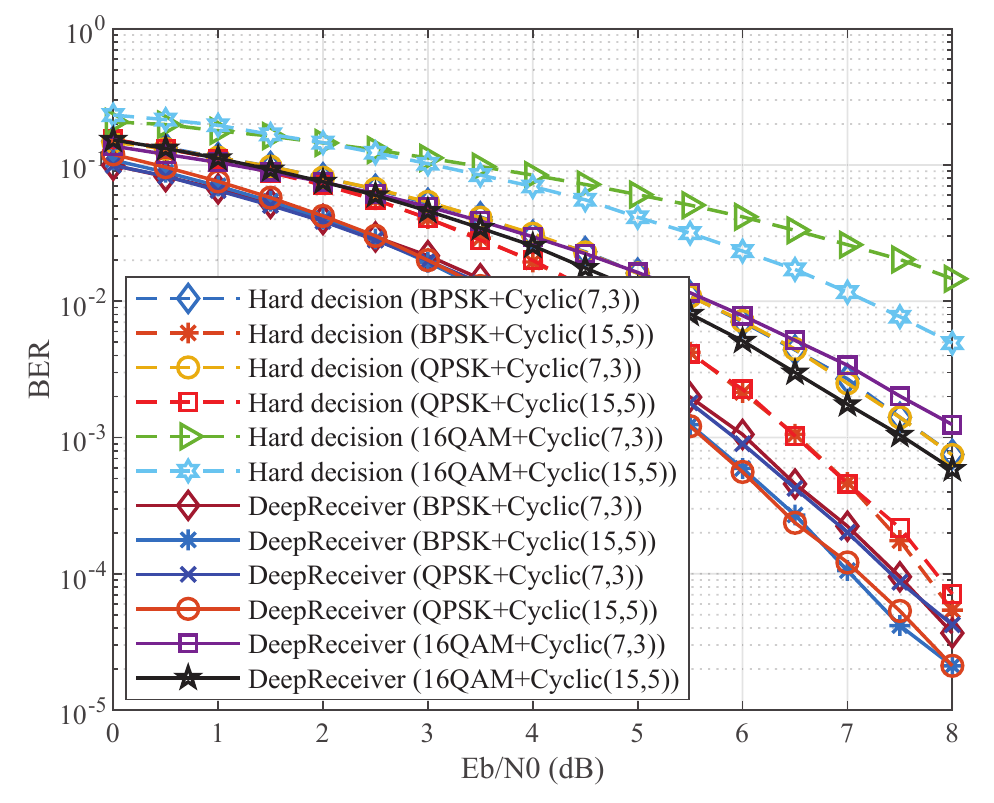}
\caption{Performance of unified blind reception for multiple MCSs. The DeepReceiver does not need to know which MCS is used while the hard decision method needs to know which MCS is used in advance.}
\label{Fig.15}
\end{figure}
In order to verify the unified information recovery ability of a single DeepReceiver model for multiple MCSs, we simulated 6 MCSs, as shown in Table I. The pulse shaping filter is a raised cosine filter with a roll-off factor of 0.5. The sampling rate of the received signal is 8 times the symbol rate. For these MCSs, the number of original information bits is 30. For 16QAM, the length of the bit stream after channel coding is not divisible by 4, we add an additional 2 bits of 0 to the bit stream. The channel noise is AWGN. The training data set contains the signals generated under these 6 MCSs. The $E_b/N_0$ ranges from 0 dB to 8 dB with an interval of 1 dB. For each MCS, the number of samples per $E_b/N_0$ is 100,000, so the total sample size of the training set is 5,400,000.

\begin{table}
\renewcommand\arraystretch{2}
  \centering
  \caption{Specification of MCSs}
  \label{Tabel.1}
  \begin{tabular}{|c|c|c|c|}
  \hline
  % after \\: \hline or \cline{col1-col2} \cline{col3-col4} ...
  MCS Index	 & Modulation & Coding & IQ Signal Length \\
  \hline
  MCS 1	& BPSK & Cyclic (7, 3) & 560 \\
  \hline
  MCS 2	& BPSK	& Cyclic (15, 5) & 720 \\
  \hline
  MCS 3	& QPSK	& Cyclic (7, 3)	& 280 \\
  \hline
  MCS 4	& QPSK	& Cyclic (15, 5) & 360 \\
  \hline
  MCS 5	& 16QAM	& Cyclic (7, 3)	& 144 \\
  \hline
  MCS 6 & 16QAM	& Cyclic (15, 5) & 184 \\
  \hline
  \end{tabular}
\end{table}

After training, we generate new samples to verify the trained DeepReceiver model. The results are shown in Fig. 15. It should be noted that the results of the hard decision methods as a comparison are obtained when the MCS of the transmitted signal is known. If the MCS is unknown, the traditional hard decision will be invalid and the resulting BER will be close to 0.5. It can be seen from the figure that the DeepReceiver can recover the original information with a very high accuracy without knowing which MCS is used for the signal, and its performance is far superior to the hard decision method. This shows that the DeepReceiver has the capability of unified receiving of multiple MCSs.

\section{Conclusions and Discussions}
In this paper, we have proposed a DeepReceiver model that uses a deep neural network to replace the receiver's entire information recovery process from the received IQ signal to the recovered information bit stream. We have proposed to use multiple binary classifiers which share the same CNN to achieve multi-bit information stream recovery. We have presented the designed 1D-Conv-DenseNet network structure to implement the DeepReceiver. We have conducted extensive simulation experiments to verify the performance of the DeepReceiver. In summary, the DeepReceiver has the following characteristics:
\begin{itemize}
  \item The DeepReceiver optimizes the overall global performance of information recovery. Simulation results have shown that the performance of the DeepReceiver can approach the ideal soft ML decision and is far superior to the hard decision method which follows serial processing of equalization, demodulation, and decoding, which verifies its end-to-end information recovery capability.
  \item The DeepReceiver can learn from the data, which better matches the non-ideal factors experienced by the communication system. The simulation results have validated the superior performance of the DeepReceiver under non-ideal conditions such as noise, RF impairment, multipath fading, and co-channel interference.
  \item The DeepReceiver can deal with co-channel interference or jamming. Simulation results have shown that the DeepReceiver has anti-interference ability under various ISRs when co-channel interference is presented, which suggests that the DeepReceiver can be served as a new anti-jamming communications method.
  \item The DeepReceiver is a blind unified receiver, which can realize the unified information recovery of multiple MCSs without knowing which MCS the received signal adopts in the inference phase (as long as the DeepReceiver has seen the signal samples of these MCSs during the training phase). Therefore, the DeepReceiver can be used for the reception of ACM, and can also be used for the unified reception of different physical layer communications protocols to facilitate the interconnection between different communications networks.

\end{itemize}

This paper mainly analyzes the performance of the DeepReceiver through simulation experiments. In the future work we will carry out experiments over the air to evaluate the performance of the DeepReceiver in the real-world environments.

%\bibliographystyle{IEEEtran}
% argument is your BibTeX string definitions and bibliography database(s)
%\bibliography{ref}

% that's all folks
\end{document}